\documentstyle[referee,epsf]{mn}
\def\gtorder{\mathrel{\raise.3ex\hbox{$>$}\mkern-14mu
                \lower0.6ex\hbox{$\sim$}}}
\def\ltorder{\mathrel{\raise.3ex\hbox{$<$}\mkern-14mu
                \lower0.6ex\hbox{$\sim$}}}
\title{Long-term optical variability properties of the Palomar-Green quasars}
\author[Giveon et al.]
       {Uriel Giveon\thanks{e-mail: giveon@wise.tau.ac.il}\thanks{web-site: http://wise-obs.tau.ac.il/$\sim$givon/DATA/}, Dan Maoz, Shai Kaspi, Hagai Netzer \\
        School of Physics and Astronomy and the Wise Observatory, 
        The Raymond and Beverly Sackler Faculty of Exact Sciences,\\
        Tel-Aviv University, Tel-Aviv 69978, Israel
\newauthor 
        and Paul S. Smith \\
        NOAO, KPNO, P.O. Box 26732, Tucson, AZ 85726-6732, USA}
\date{Accepted 1999 February 16,
      Received 1998 September 9}

\pagerange{\pageref{firstpage}--\pageref{lastpage}}
\pubyear{1998}

\begin{document}
\maketitle

\label{firstpage}

\begin{abstract}
We present results from a monitoring program of 42 quasars from the Palomar-Green sample. The objects were observed for seven years at the Wise Observatory, as part of a long term effort to monitor AGN of various types. This is the most extensive program of its kind carried out to date on a well-defined optically-selected quasar sample. The typical sampling interval is $\sim 40$ days. One third of the quasars were observed at $\sim 60$ epochs and the rest at $\sim 30$ epochs in two bands ($B$ and $R$) with photometric accuracy of $\sim 0.01$ mag.
We present lightcurves for all of the sources and discuss the sample variability properties.
All of the quasars in the sample varied during the campaign with intrinsic rms amplitudes of $5\%<\sigma_B<34\%$ and $4\%<\sigma_R<26\%$. The rms amplitude and colour for the entire sample are $\sigma_B=14\%$, $\sigma_R=12\%$, and $\sigma_{B-R}=5\%$. On time scales of 100 to 1000 days the power spectra of the sources have a power-law shape, $P_{\nu}\propto \nu^{-\gamma}$, with $\gamma\approx 2.0$ and a spread $\ltorder 0.6$.
At least half of the quasars, particularly those that are most variable, become bluer when they brighten, and the rest do not show this behaviour. We quantify this phenomenon, which has been observed previously mainly in Seyfert galaxies.
The quasars which are most variable tend also to exhibit asymmetry in their variations, in the sense that the brightening phases last longer than the fading phases.
We have searched for correlations between the measured variability properties and other parameters of the quasars, such as luminosity, redshift, radio loudness, and X-ray slope. We find several new correlations, and reproduce some of the correlations reported by previous studies. Among them are an anticorrelation of variability amplitude with luminosity, a trend of the autocorrelation time scale with luminosity, and an increase in variability amplitude with H$\beta$ equivalent width.
However, all of these trends have a large scatter despite the low observational uncertainties.
\end{abstract}



\section{Introduction}

\label{introduction}

Continuum variability is a common property of active galactic nuclei (AGN), and is observed in all accessible spectral bands, from gamma-rays to radio. AGN lightcurves show structure over a wide range of time scales and amplitudes. It is now well established that there are at least two variability classes of AGN. The first includes BL Lac objects and optically violent variable (OVV) quasars, whose variability is characterized by short time scales and large amplitudes -- properties which are usually attributed to relativistic beaming (e.g., Bregman 1990). The second class, which includes the majority of AGN ($\sim 90 \%$), is the smaller-amplitude variables, which show variations of up to tens of percents in amplitude on time scales of weeks to years. The physical origin of their variability may be completely different, e.g., instabilities in accretion disks around massive black holes (Siemiginowska \& Elvis 1997; Kawaguchi et al. 1998).

Among the many studies of AGN variability in the various spectral bands, many workers have attempted to quantify the optical variability of quasars, and have discussed its correlations with redshift, luminosity, and spectral properties.
Table \ref{former_works} summarizes the observations and the results of some major studies of quasar variability.
\begin{figure*}
\includegraphics{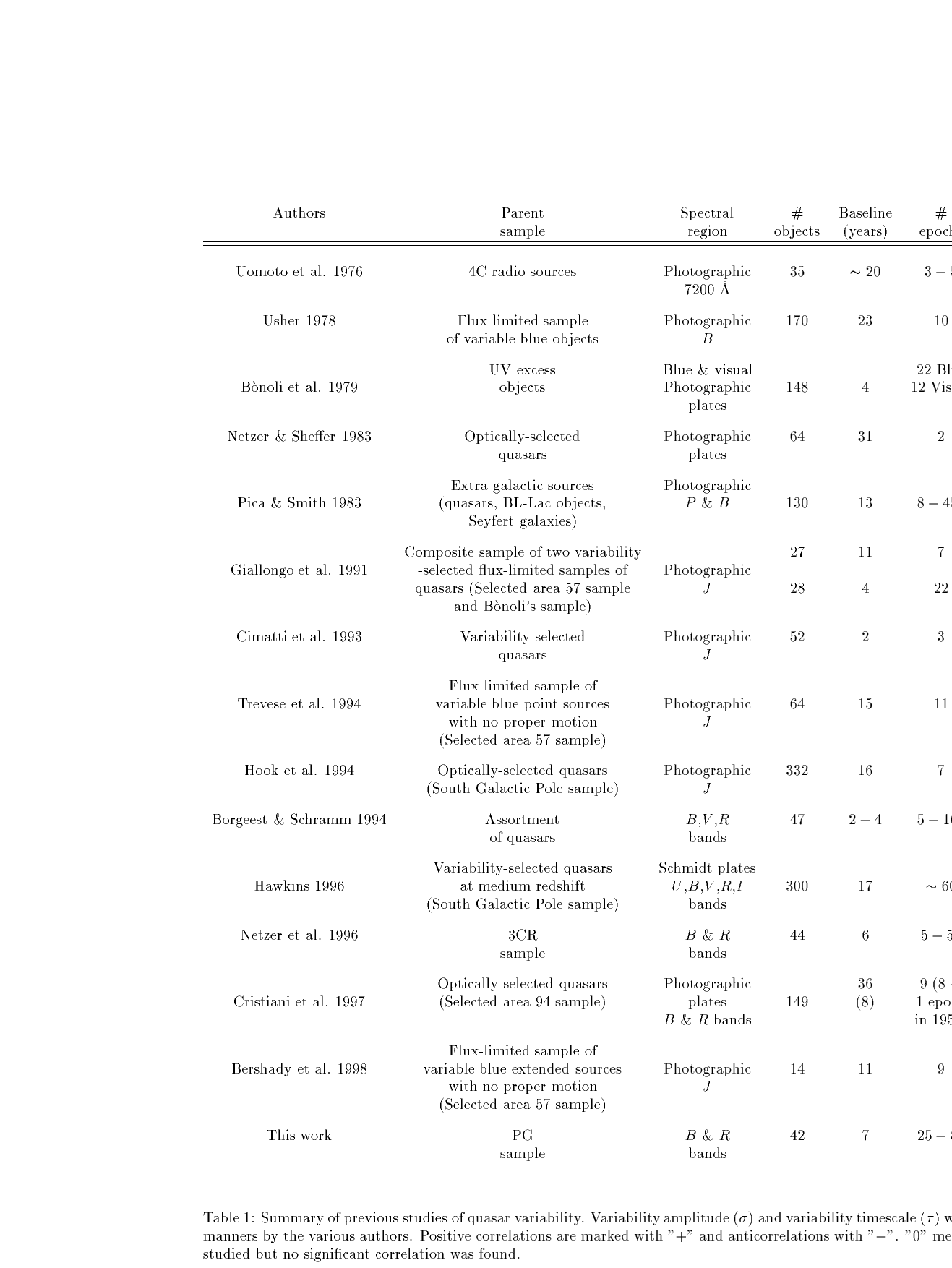}
\vspace{25cm}
\end{figure*}
\begin{table*}
\label{former_works}
\end{table*}
\addtocounter{table}{1}We list only the most recent paper from each program.
 
The motivation for this extensive research has been the hope to constrain the nature of the central engines of quasars using variability data. Unfortunately, even basic results, such as whether luminosity and variability amplitude of quasars are correlated, are discordant among different studies. This suggests that sample selection, temporal sampling, and measurement errors may be affecting the conclusions.
In early studies, samples were often chosen arbitrarily and concentrated on bright or ``interesting'' objects, that showed large variations.
Such samples may be unrepresentative of the population of quasars as a whole, and the results may simply reflect the selection criteria.
Even well-defined samples may produce spurious results due to irregular sampling. The sampling of most previous monitoring was sparse, making the variability properties sensitive to a few data points. The lightcurves were usually also unevenly sampled, complicating most time series analyses.

We are engaged in several long-term quasar monitoring programs at Wise Observatory that attempt to overcome some of these problems. Netzer et al. (1996) described a 3-6 year monitoring program of radio-loud quasars.
The present paper presents the results of a program aimed at an optically selected quasar sample. We have monitored a well-defined subset of the Palomar-Green QSO sample (PG; Schmidt and Green 1983) for seven years. 
The PG sample is one of the best-studied optically-selected quasar samples. Its properties have been quantified at many wavelengths: radio - Kellermann et al. (1989), IR - Neugebauer et al. (1987), optical - Miller et al. (1992), and X-ray - Laor et al. (1997).
The optical emission-lines and their variability in the PG sample were studied by Boroson \& Green (1992), Maoz et al. (1994), and Kaspi et al. (1996), and the sample's X-ray variability properties were studied by Fiore et al. (1998).
The polarization properties of the PG sample were studied by Berriman et al. (1990).

Comparison between the various studies listed in Table \ref{former_works} needs to be done with care, because the programs sample different regions of the absolute magnitude-redshift plane.
In the present work, most of the sample are low-luminosity (and low-redshift) quasars (see \S \ref{sample}). By the commonly-adopted absolute-magnitude criterion for quasars, $M_B < -22$, a minority of the sample may be considered Seyfert galaxies.
Our sample and the samples of B\`{o}noli et al. (1979), Giallongo et al. (1991), Cimatti et al. (1993), and Trevese et al. (1994) consist of quasars of similar absolute magnitudes ($-28<M_B<-21$, see Figure \ref{maghistograms}), but the redshift ranges in those studies are broader.
Variability studies of {\it bona-fide\/} Seyfert galaxies have mainly focused on individual objects (e.g., Clavel et al. 1991), chosen by virtue of their high variability. The work of Bershady et al. (1998) is an attempt to quantify the variability of Seyfert galaxies in general.
In terms of observational parameters, the sampling rate of our program is much higher and more even than that of past programs.
The high accuracy attainable with differential CCD photometry allows us to probe smaller amplitude variations.

In \S \ref{observations_reduction} we describe the sample, the observations, and the reduction. In \S \ref{analysis} we describe the time-series and correlation analyses. In \S \ref{conclusions} we summarize our conclusions. 

\section{Observations}

\label{observations_reduction}

\subsection{The Sample}

\label{sample}

Our goal is to investigate the long-term variability properties of a particular sample of optically selected quasars and their connection to other quasar properties. The PG sample is well suited for this purpose, being a statistically complete sample (however, see Goldschmidt et al. 1992; K\"{o}hler et al. 1997) of UV-excess quasars. We chose from the PG sample all nearby (redshift $z<0.4$), bright ($B<16$ mag according to Schmidt \& Green 1983) quasars (absolute $B$ magnitude $M_B< -22$ mag, assuming a Hubble constant $H_0=70 \ {\rm km\ s}^{-1}{\rm Mpc}^{-1}$, a deceleration parameter $q_0=0.2$, and a power-law-shaped continuum, $F_{\nu}\propto \nu^{-\alpha}$, with spectral index of $\alpha=0.5$ for the K-correction; see Schmidt \& Green 1983) with northern declinations. We also required that there be a star within 4 arcminutes of the quasar with comparable brightness. The redshift criterion was applied so that the main optical emission lines (H$\alpha$ and H$\beta$) remain within the optical range for the purposes of a spectroscopic study (Maoz et al. 1994; Kaspi et al. 1996) carried out in parallel to the present work. The properties of the 42 PG quasars passing these criteria are summarized in Table 2 in the following format:
\begin{itemize}
\item[]{\it Column (1)\/}---Right ascension (1950).
\item[]{\it Column (2)\/}---Declination (1950).
\item[]{\it Column (3)\/}---Redshift.
\item[]{\it Column (4)\/}---Number of observing epochs.
\item[]{\it Column (5)\/}---Total observing period, in days.
\item[]{\it Column (6)\/}---Median sampling interval, in days.
\item[]{\it Columns (7)-(10)\/}---Median apparent and absolute magnitudes in the $B$ and the $R$ bands based on our measurements. The absolute magnitudes were computed assuming a spectral index,
\begin{equation}
\alpha_{B-R}={{med(B-R)-(B_0-R_0)}\over{2.5\log{{\lambda_R}\over{\lambda_B}}}}+2,\label{spectal_slope}
\end{equation}
where $B_0$ and $R_0$ are the zero-points of the Johnson-Cousins magnitude system (Allen 1983), and $\lambda_B=4400$ \AA\ and $\lambda_R=6400$ \AA\ are the central wavelengths of the $B$ and the $R$ bands, respectively.
Using our measured colours and time-averaged magnitudes, the sample distribution of absolute magnitudes is somewhat changed from that used to define the sample, and no longer has a sharp cutoff. Figure \ref{maghistograms} shows the original distribution of absolute magnitudes and the updated one.
\begin{figure}
\centerline{\epsfxsize=3.8in\epsfbox{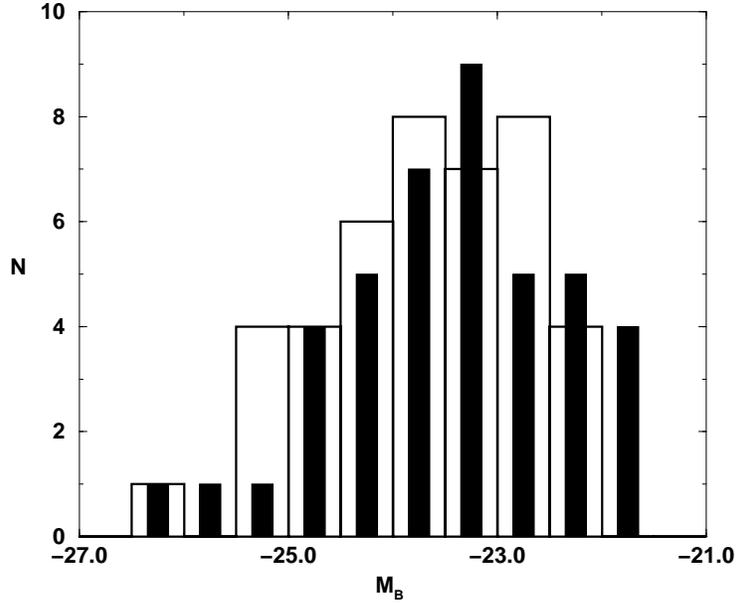}}
\caption{Distributions of absolute $B$ magnitudes of the sample. The original distribution used for sample selection is represented by empty bars, and the distribution based on our measurements is represented by filled bars.}
\label{maghistograms}
\end{figure}
\item[]{\it Column (11)\/}---Radio (5 GHz)-to-optical ($B$ band) flux ratio (Kellerman et al. 1989).
\item[]{\it Column (12)\/}---Radio power at 5 GHz in ${\rm W\ Hz}^{-1}$ (Kellerman et al. 1989).
\item[]{\it Column (13)\/}---H$\beta$ full width at half maximum (FWHM), in ${\rm km\ s}^{-1}$ (Boroson \& Green 1992).
\item[]{\it Column (14)\/}---ROSAT-PSPC X-ray spectral index $\alpha_x$ ($F_{\nu}\propto \nu^{-\alpha_x}$, Laor et al. 1997).
\item[]{\it Column (15)\/}---[OIII] $\lambda 5007$ equivalent width, in \AA\ (EW, Boroson \& Green 1992).
\item[]{\it Column (16)\/}---FeII EW, in \AA\ (Boroson \& Green 1992).
\item[]{\it Column (17)\/}---[OIII] $\lambda 5007$ to H$\beta$ peak intensities ratio (Boroson \& Green 1992).
\item[]{\it Column (18)\/}---H$\beta$ EW, in \AA\ (Boroson \& Green 1992).
\item[]{\it Column (19)\/}---HeII $\lambda 4686$ EW, in \AA\ (Boroson \& Green 1992).
\item[]{\it Column (20)\/}---Optical (2500 \AA)-to-X-ray (2 keV) spectral index, $\alpha_{ox}$. We use our optical data to derive the 2500 \AA\ luminosity. X-ray data are taken from Tananbaum et al. (1986), Kriss (1988), and Laor et al. (1997). Laor et al.'s, based on ROSAT measurements, are the most reliable, and were preferred in case of disagreement. 

We have also studied the IR spectral indices (Neugebauer et al. 1987) and the polarization levels (Berriman et al. 1990) of the PG quasars. These parameters do not show significant correlations with any variability parameter we have measured (see \S \ref{discussion}), so for brevity's sake they are not listed in Table 2.
\end{itemize}
\begin{figure*}
\includegraphics{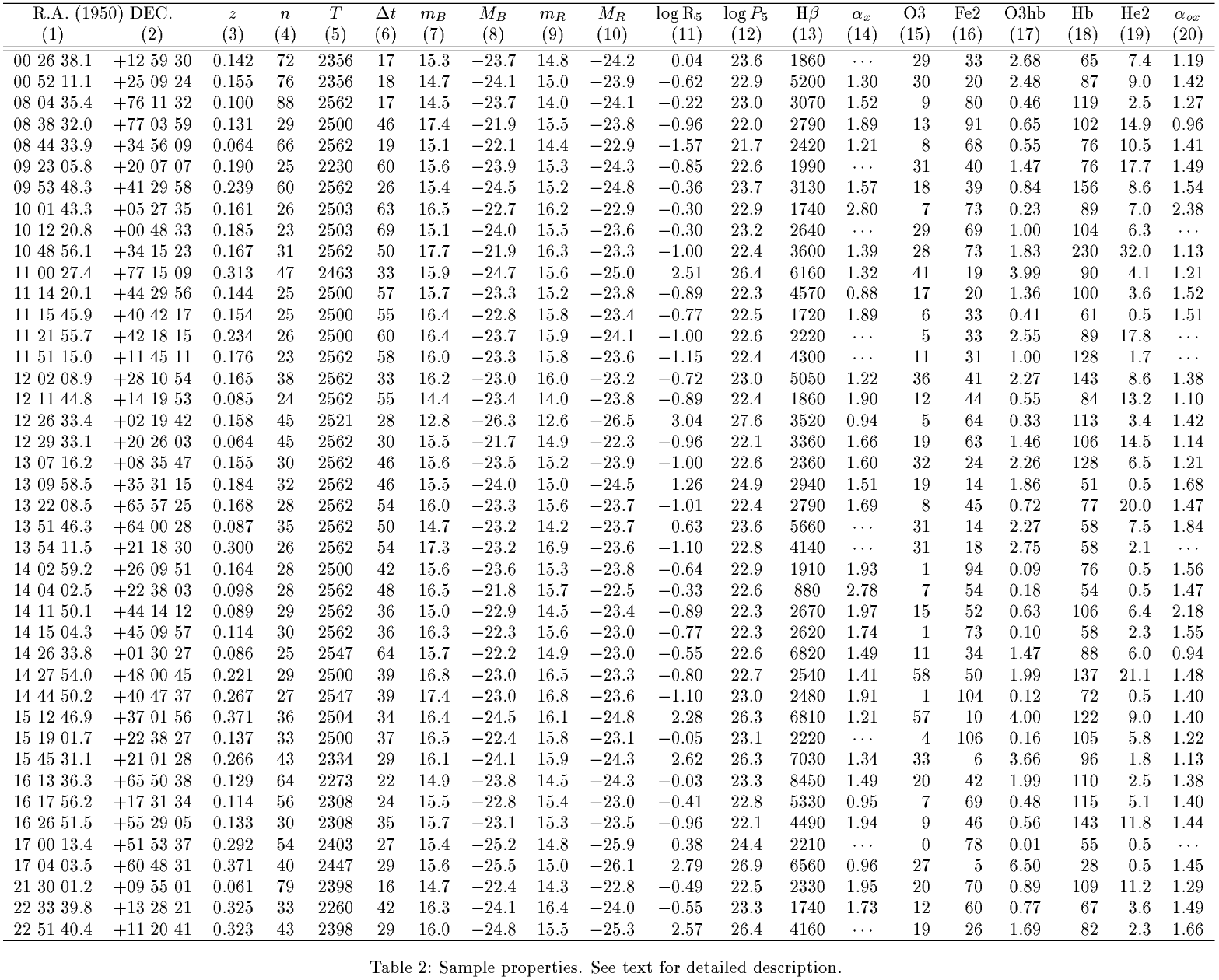}
\vspace{25cm}
\end{figure*}
\begin{table*}
\label{data}
\end{table*}
\addtocounter{table}{1}

\subsection{Observations and Data Reduction}

All objects were monitored for 7 years (1991 March through 1998 March) in the Johnson-Cousins $B$ and $R$ bands with the Wise Observatory 1 m telescope in Mitzpe-Ramon, Israel. The sampling properties (including the number of observing epochs, total period of observations, and the median sampling interval) of each object are listed in Table 2. The typical (median over the entire sample) sampling interval is 39 days.

Observations from 1991 March through 1994 January were obtained using an RCA  $320\times 520$-pixel thinned CCD.
Most of the data (1994 February through 1998 March) were obtained using a Tektronix $1024\times 1024$-pixel back-illuminated CCD, with typical integration times of 2-3 minutes in $R$ and 3-4 minutes in B. 
During an intermediate period (1993 April through June), a TI $1000\times 1024$-pixel CCD was used.
Spectrophotometric monitoring of twenty-eight objects from our sample was carried out during the same period at Wise Observatory and at Steward Observatory (Maoz et al. 1994; Kaspi et al. 1996; Kaspi et al. 1999). Flux calibrated spectra for thirteen of the quasars are presently available. We multiplied those spectra with the Johnson-Cousins $B$ and $R$ transmission curves and integrated them over wavelength to derive $B$ and $R$ magnitudes, and thereby improved the temporal sampling of these objects. 

Owing to the large volume of data ($\sim 3500$ CCD frames), reduction was carried out by specially written automatic routines that produced lightcurves from raw images with minimal interaction.
Photometry was obtained with the point spread function (PSF) fitting routines in the DAOPHOT (Stetson 1987) package within IRAF\footnote[2]{IRAF (Image Reduction and Analysis Facility) is distributed by the National Optical Astronomy Observatories, which are operated by AURA, Inc., under cooperative agreement with the National
Science Foundation.} applied to the bias-subtracted and flat-field-corrected frames.
This produced frame-dependent instrumental magnitudes for each quasars and for $5-10$ reference stars in its field.
One image of each quasar was used as a reference image. For each reference star, the difference between its magnitude in a given image and its magnitude in the reference image was added to the quasar magnitude in the image to obtain a quasar lightcurve relative to each reference star. These lightcurves where then averaged to obtain the final quasar lightcurve.
Reference stars whose magnitude difference deviated by more than $3\sigma$ from the mean magnitude difference of the other reference stars at a certain epoch (presumably due to systematics such as bad pixels, cosmic rays, or variations of the reference stars) were excluded. Reference star measurements with photometric errors larger than 0.1 mag were also excluded.
Photometric standards (Landolt 1992) were observed on several clear nights and were used to bring the lightcurves to an absolute scale.

\begin{figure*}
\caption{Quasar lightcurves: Filled symbols are measurements in the $R$ band and empty symbols are measurements in the $B$ band. Circles are photometric measurements and squares are spectrophotometric measurements. In the cases indicated, the $B$ band lightcurves were vertically shifted as marked to enable uniform scale for all objects. Only the panel for PG1626+554 has a different scale.}
\label{lightcurves}
\centerline{\epsfxsize=7.5in\epsfbox{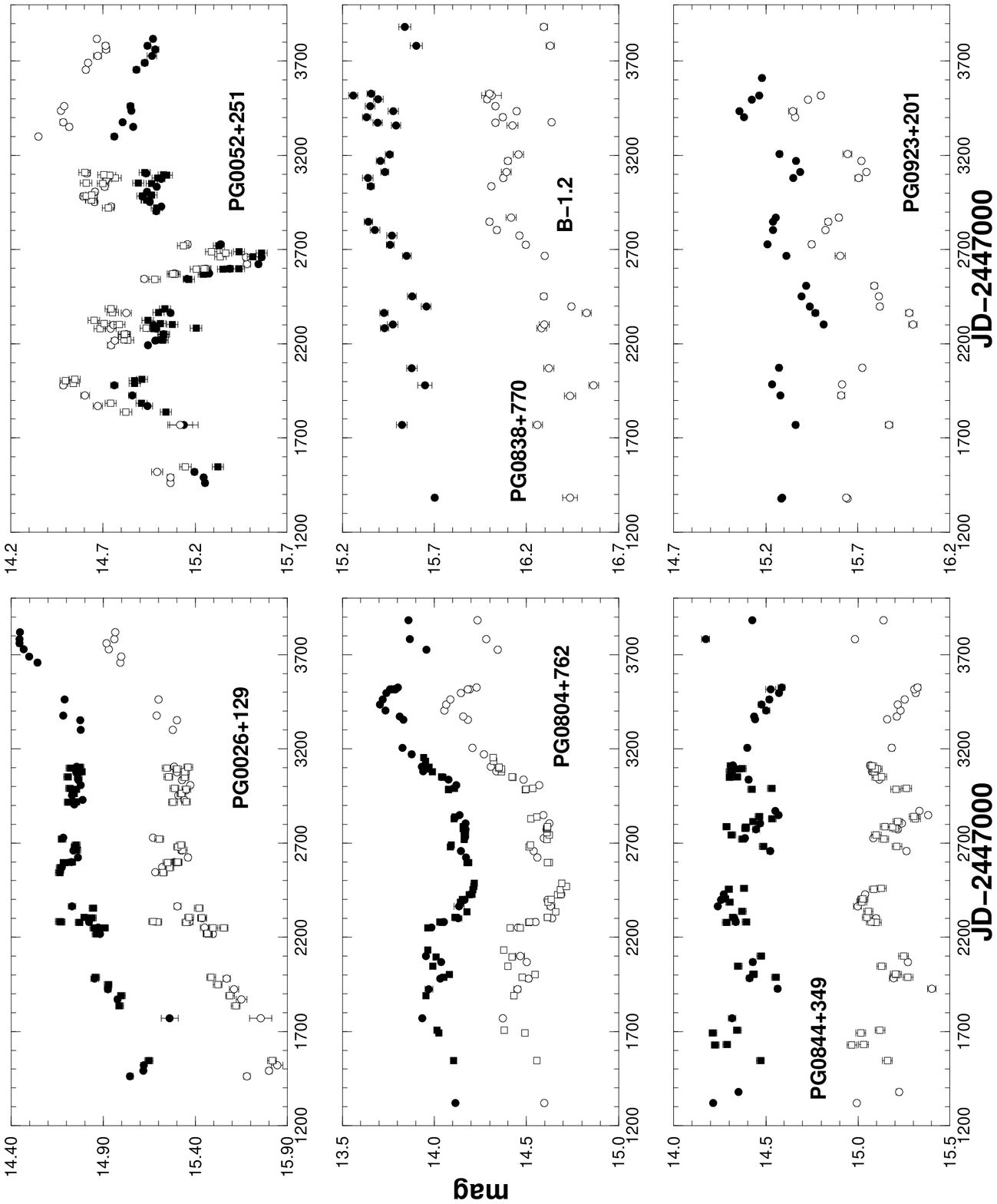}}
\end{figure*}

Lightcurves at $B$ and $R$ for each object are shown in Figure \ref{lightcurves}. A machine-readable version of these data is available upon request from the authors.
\begin{figure*}
\centerline{\epsfxsize=7.5in\epsfbox{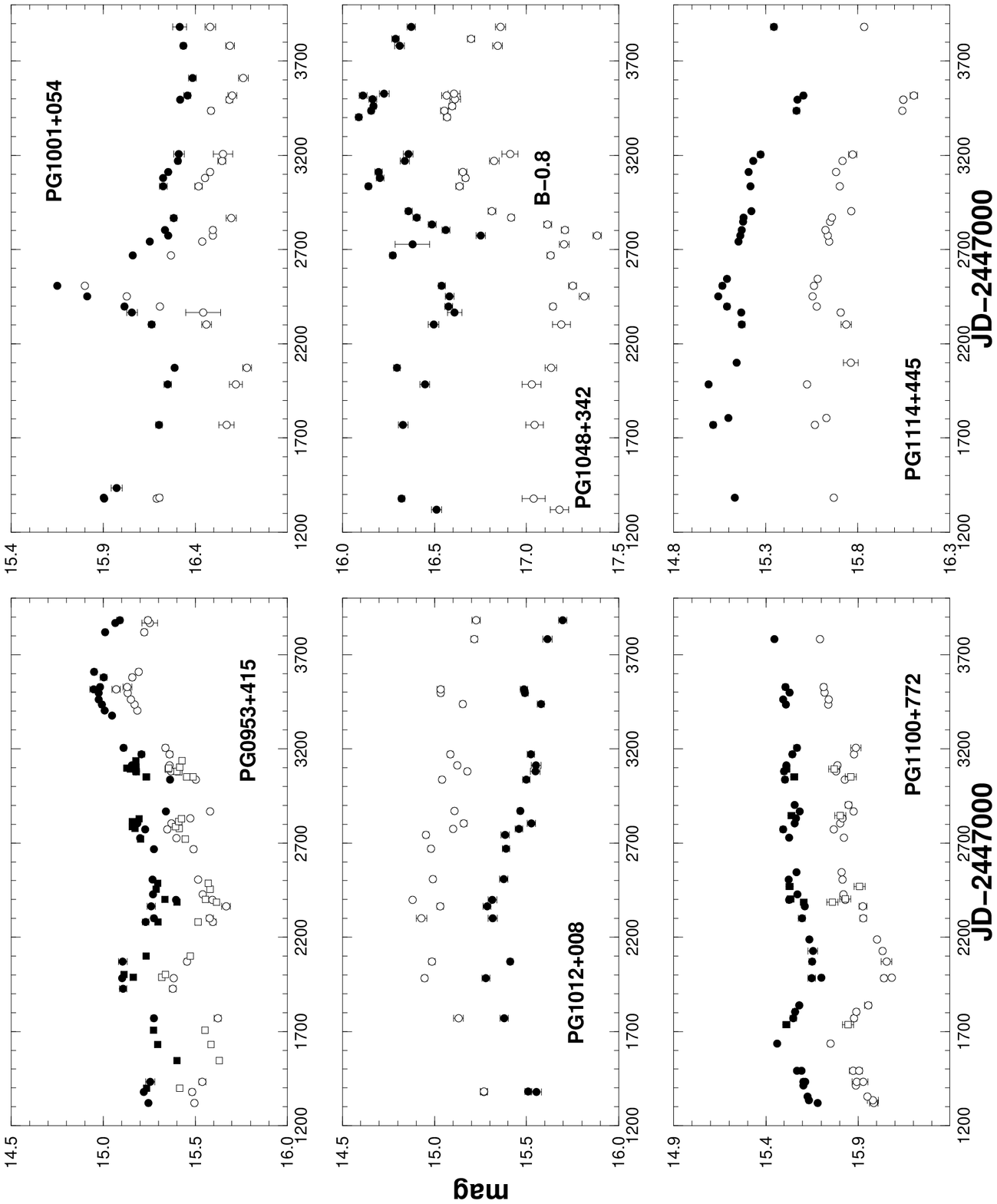}}
\end{figure*}
\begin{figure*}
\centerline{\epsfxsize=7.5in\epsfbox{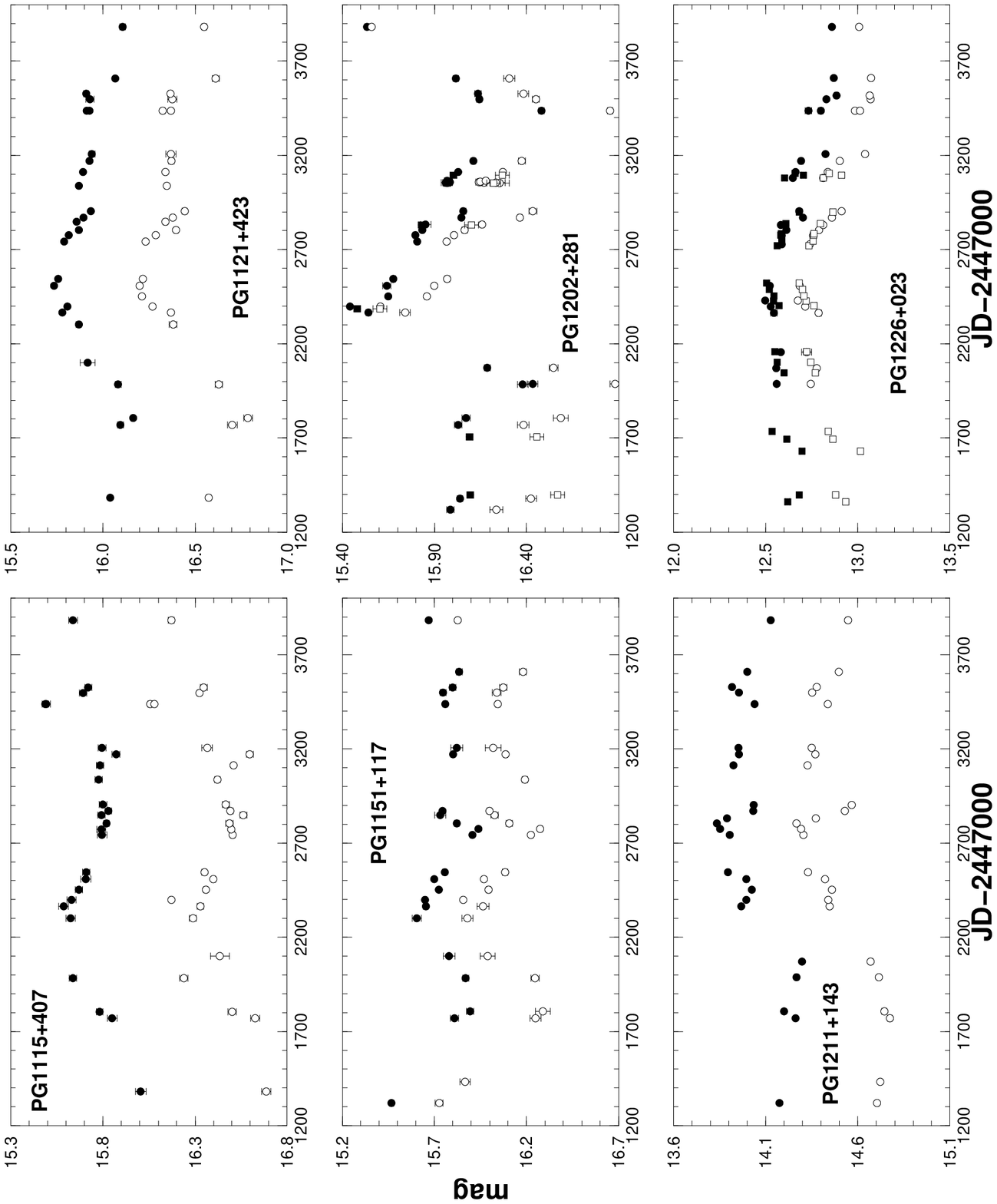}}
\end{figure*}
\begin{figure*}
\centerline{\epsfxsize=7.5in\epsfbox{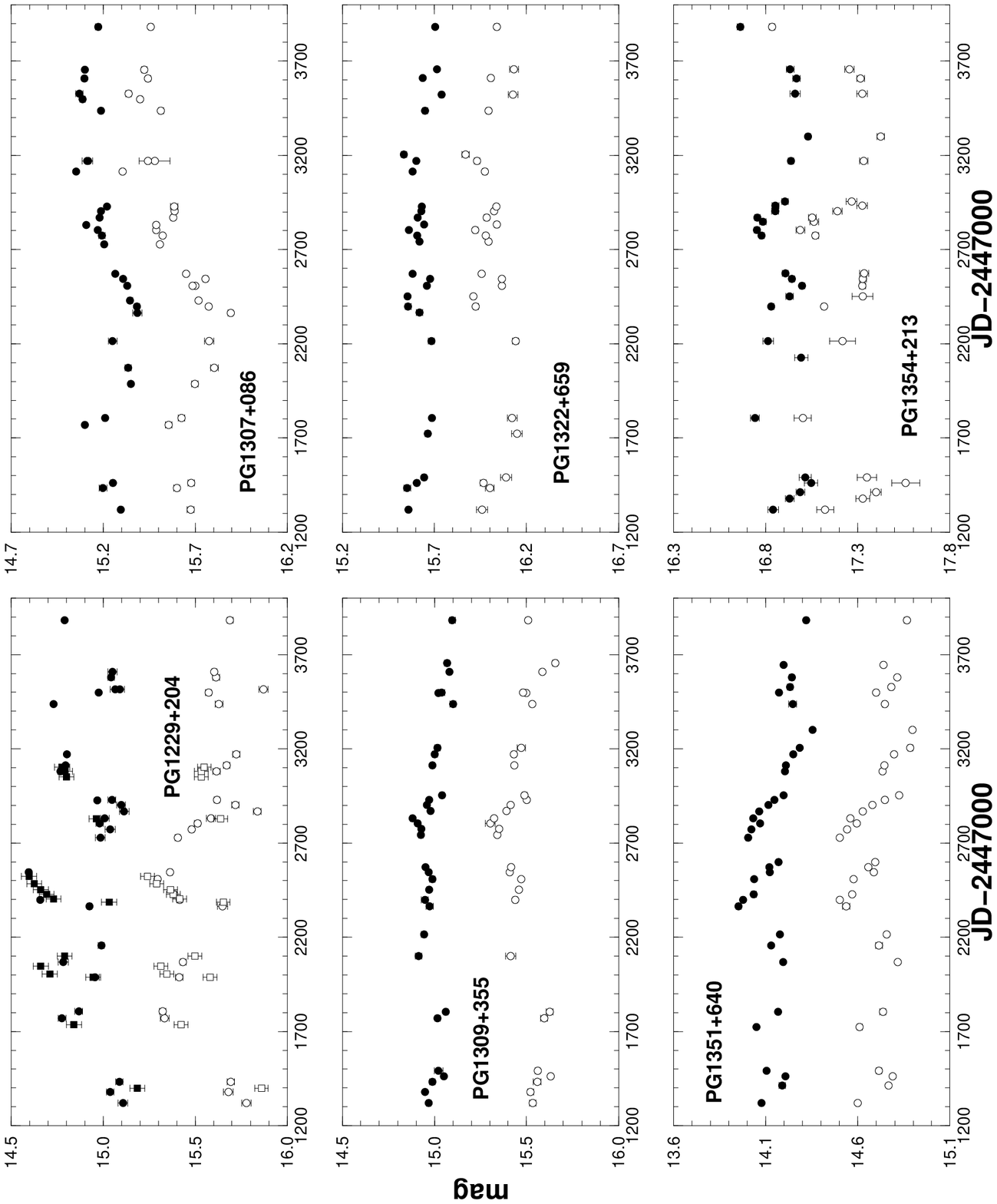}}
\end{figure*}
\begin{figure*}
\centerline{\epsfxsize=7.5in\epsfbox{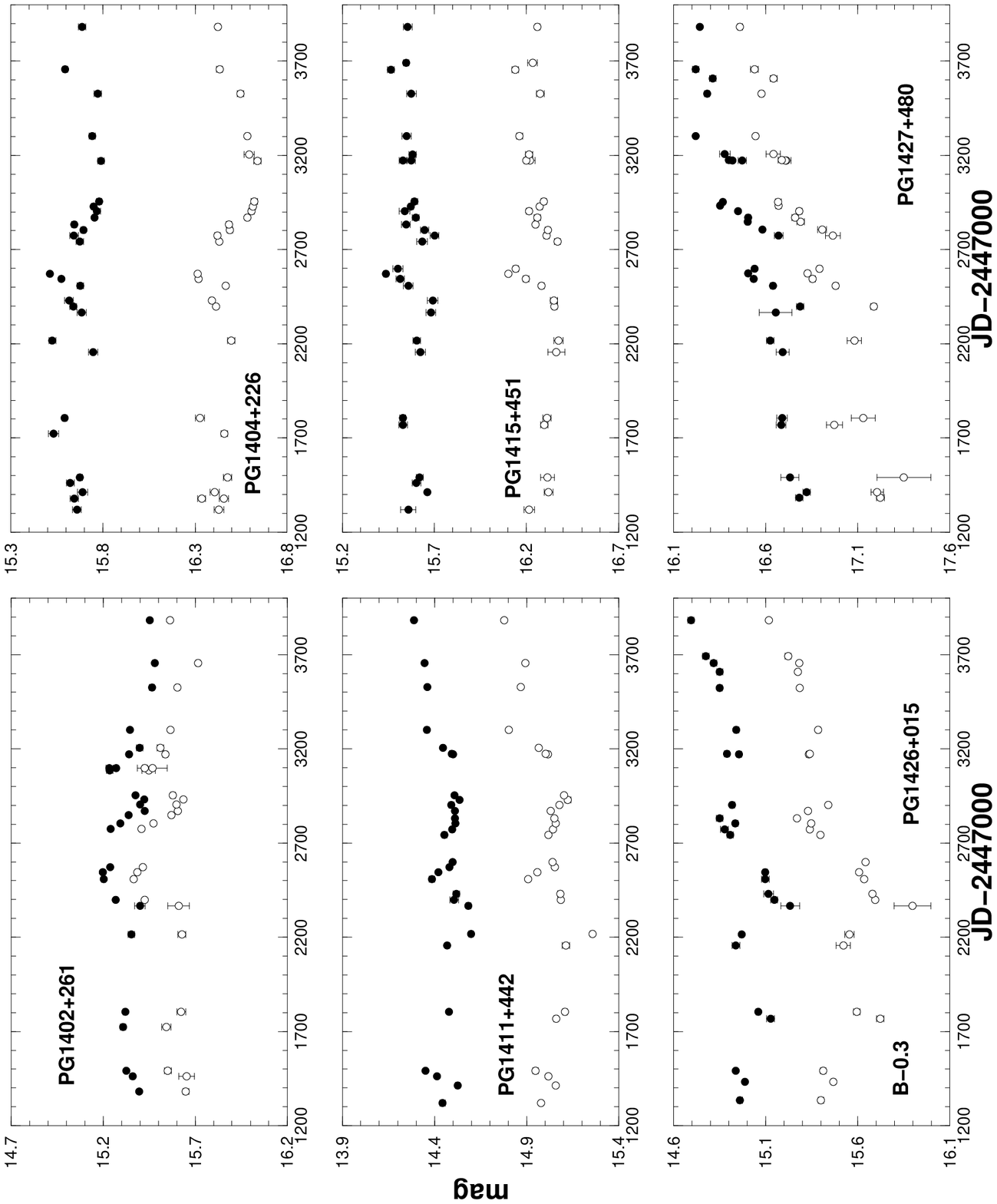}}
\end{figure*}
\begin{figure*}
\centerline{\epsfxsize=7.5in\epsfbox{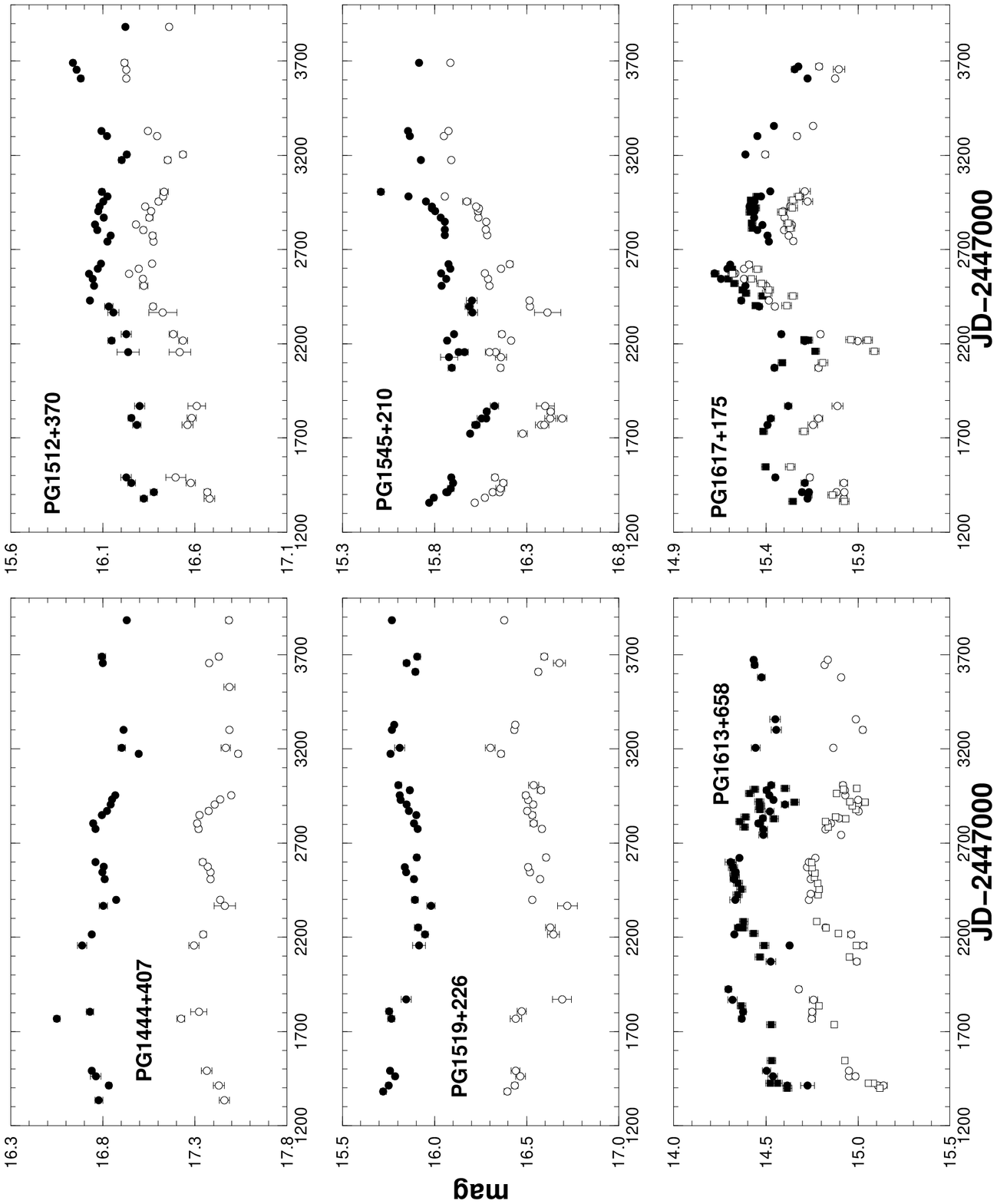}}
\end{figure*}
\begin{figure*}
\centerline{\epsfxsize=7.5in\epsfbox{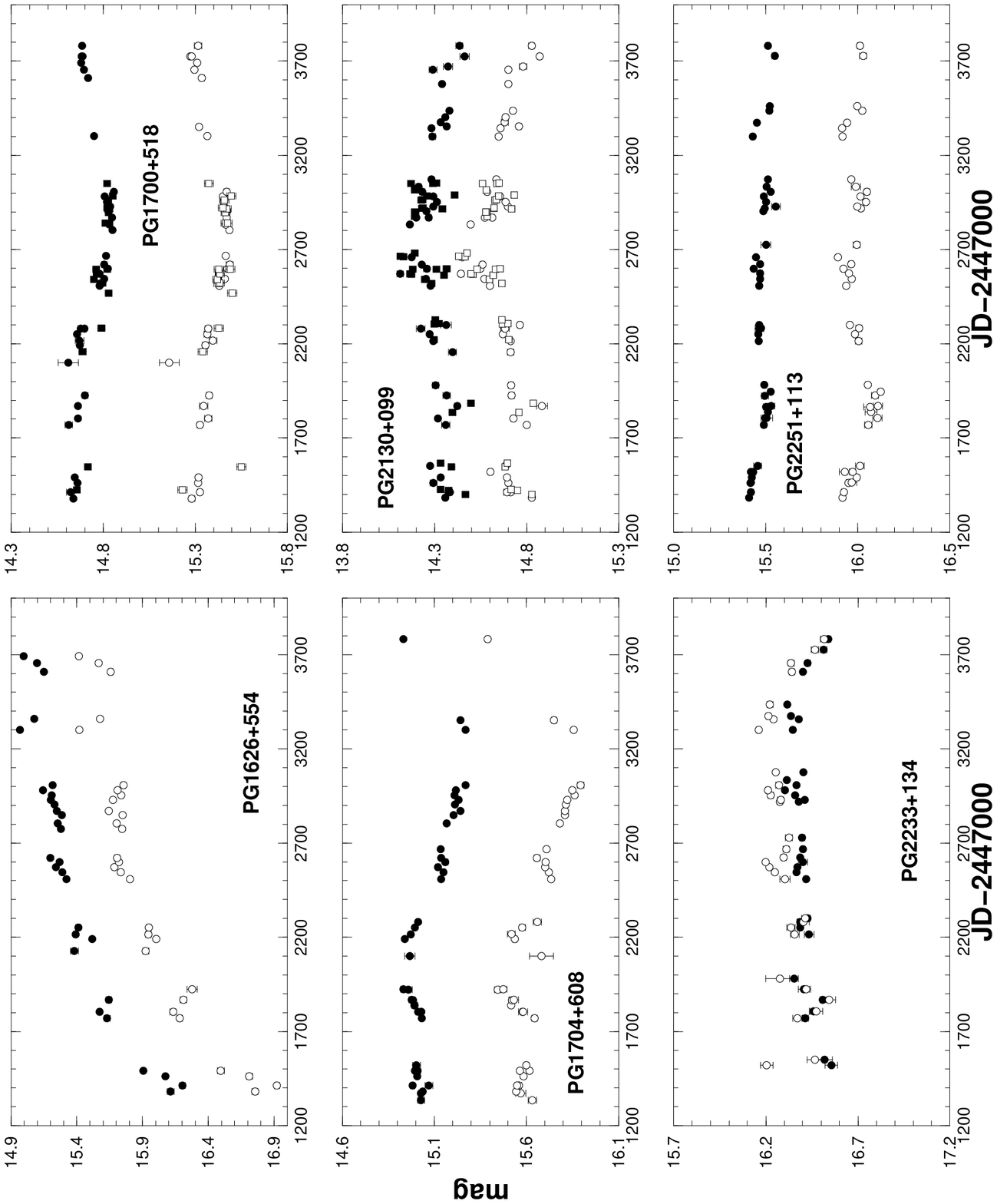}}
\end{figure*}The uncertainty in each point is a combination in quadrature of the uncertainty of the PSF fitting error and the $1\sigma$ scatter of the magnitude differences of the reference stars between that epoch and the reference epoch. Uncertainties are $\sim 0.01$ mag in the $R$ band and somewhat higher ($\sim 0.02$ mag) in the $B$ band, due to lower CCD sensitivity at the blue wavelengths.
The uncertainties are dominated by the PSF fitting errors. Aperture photometry with large fixed apertures did not improve the accuracy. The uncertainty in the absolute photometric scale, which is typically 0.08 mag, is not included in the error bars.

\section{Analysis}

\label{analysis}

\subsection{Ensemble Variability Amplitude}

\label{ensemble}

\begin{figure}
\centerline{\epsfxsize=3.8in\epsfbox{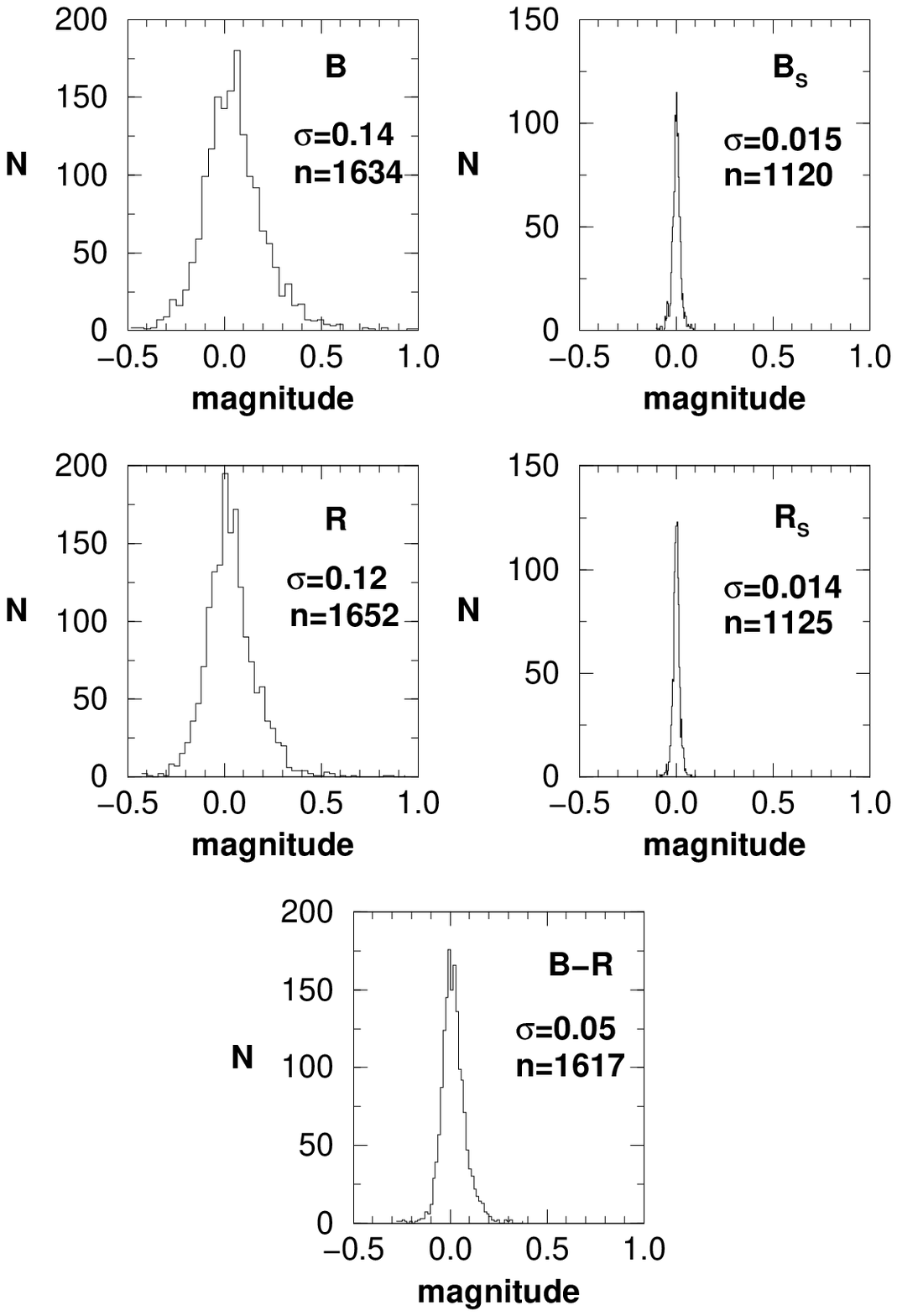}}
\caption{Distributions of the brightness deviations about the mean lightcurve levels of the $B$, $R$, and $B-R$ measurements of the entire sample. $B_S$ and $R_S$ are the brightness deviation distributions of stars in the quasars' fields.}
\label{histograms}
\end{figure}

Figure \ref{histograms} shows histograms of the brightness deviations about the mean levels in $B$, $R$, and $B-R$, along with their $1\sigma$ scatter and the number of points contributing to the histogram.
Also shown are the brightness deviation distributions of stars (one for each quasar field), $B_S$ and $R_S$. The stellar distributions do not include spectrophotometric measurements (there is only one star per quasar in those measurements) and hence have fewer points.
The standard deviations of the stellar distributions provide an independent estimate of the systematic error over all epochs for the entire sample: $\sigma_{BS}=0.015$ mag and $\sigma_{RS}=0.014$ mag.
Between February and November 1997 the telescope suffered from scattered-light problems as a result of a change in baffling. Data from this period have larger systematic errors ($0.03-0.04$ mag) due to the resulting flat-fielding problems.
The following values can be regarded as the typical intrinsic variability amplitudes of the entire sample:
\begin{equation}
\hat{\sigma}_B=\sqrt{\sigma_B^2-\sigma_{BS}^2}=0.14\ {\rm mag}
\end{equation}
\begin{equation}
\hat{\sigma}_R=\sqrt{\sigma_R^2-\sigma_{RS}^2}=0.12\ {\rm mag}
\end{equation}
\begin{equation}
\hat{\sigma}_{B-R}=\sqrt{\sigma_{B-R}^2-\sigma_{BS}^2-\sigma_{RS}^2}=0.05\ {\rm mag}
\end{equation}
In principle, the constant contribution from the host galaxies of the quasars may dilute their intrinsic variations. 
Bahcall et al. (1997) observed the host galaxies of ten quasars that are in our sample with the {\it Hubble Space Telescope\/}.
From their measurements we estimate the galaxy contribution to be $\sim 5\%$ within our PSF fitting diameter of $\ltorder 3''$. The intrinsic quasar variations are therefore $\sim 1.05$ times larger than the numbers above.
Nevertheless, the PG quasars in Bahcall et al. (1997) do not cover the entire range in redshift and luminosity of our sample, and so may not be completely representative. The host galaxies may also slightly change the colour of the quasars (see \S \ref{discussion}).

Another factor that can bias the variability parameters is the emission-line contribution to the broad-band fluxes. In the PG sample, the typical emission-line contributions to the $B$ and $R$ fluxes are $5-10\%$ and $10-15\%$, respectively. This includes both the constant narrow lines and the variable broad lines. We further discuss this issue in \S \ref{discussion}

\subsection{Statistical Measures}

We define below the statistics we use to estimate the variability amplitude and time scale, and the method by which we test for the correlations between these parameters and other quasar parameters.

{\bf Variability Amplitude -}
We have estimated the variability amplitude of each object in various ways. The standard deviation was calculated for each lightcurve, weighting each data point by its inverse variance. The maximal peak-to-peak difference between any two points in the lightcurve is another conventional estimator used by many authors (e.g., Netzer et al. 1996).
Other meaningful measures are the variability mean and median, defined as the mean and median of all possible magnitude differences of a lightcurve. The median is a better estimator in terms of its lower sensitivity to outliers (Hook et al. 1994; Netzer et al. 1996).
We have also calculated separately the medians, $median(\Delta m_+)$ and $median(\Delta m_-)$, of the positive and the negative differences for each lightcurve, in order to test for asymmetries in the lightcurves. We define the net difference of the lightcurve by:
\begin{equation}
\delta m=median(\Delta m_+)-median(|\Delta m_-|).
\label{netdiff}
\end{equation}
The standard deviations and the mean, median, maximal, and net differences of the lightcurves are listed in Table 3.

{\bf Variability Gradient -}
A simple estimate of variability is the variability gradient (Borgeest \& Schramm 1994), defined as the magnitude change per unit time.
Using all pairs of adjacent data points for each lightcurve (including the $B-R$ lightcurves), we have calculated the weighted mean, median, and maximal gradient, in units of magnitude per year, in the quasars' rest-frame, and list them in Table 3.
For the maximal gradient we took only pairs of points that are more than one day apart, since the combination of the small flux fluctuations (due to photometric uncertainties) over a few hours can translate into unrealistically large values in magnitudes per year.
We note that the variability gradient is a crude estimator. Two completely different lightcurves in terms of typical time scale, such as a periodic low-amplitude time series and a time series with a single high-amplitude flare, can give the same mean gradient.

{\bf Power Spectrum Analysis -}
A time scale can also be measured, in principle, from features in a Fourier power spectrum. Previous observations of individual AGN, especially in the X-ray range, have yielded a featureless, power-law-shaped power spectrum, $P(\nu)\propto \nu^{-\gamma}$, on time scales ranging from minutes to weeks (e.g., Green et al. 1993; Paltani et al. 1997; Fiore et al. 1998). Recent study of NGC 3516 by Edelson \& Nandra (1998) have shown that its power spectrum exhibits a break on time scale of $\sim 30$ days. We have performed a power spectrum analysis on the mean-subtracted lightcurves of the individual objects.

There are algorithms (e.g., the Lomb-Scargle algorithm, Scargle 1989) designed for power spectrum analysis of unevenly-spaced data, whose basic approach is to place zeroes on an evenly-sampled grid at all points in the mean-subtracted lightcurve where there are no data. This approach works well when one is searching for periodicities. However, when dealing with noise-like spectra, such as those characterizing AGN, the above approach can distort the true slope of the power spectrum.
This occurs because a lightcurve with zeroes where there are no data will be much ``bluer'' (i.e., with many short, large spikes, producing relatively too much power on short time scales) than it would be had it been  evenly sampled. Stated differently, the long time scales do not get enough weighting in the Fourier sum, because the slow harmonics get multiplied mostly by zeroes. One can, alternatively, interpolate the lightcurve to an evenly-sampled grid, but this involves inventing artificially-smoothed variations in the observational gaps, with the effect of attenuating the high frequencies in the power spectrum. 
As a compromise between these two extremes (no interpolation and full interpolation) we have used a scheme (T. Mazeh 1998, private communication)where interpolated data are used only when the interval between the real and interpolated data points is smaller than a set fraction of the Fourier time scale being calculated. This appears to work well, in that only the desirable aspects of each approach are kept. Specifically, for every frequency $\nu$ whose power is calculated, a linearly interpolated point is introduced in the lightcurve only if the time-difference, $\Delta t$, between the interpolated point and the closest real data point is smaller than $1/2\nu$. Otherwise, the flux is set to zero at this epoch in the evenly-sampled grid.

The two lowest frequency points of each power spectrum were ignored, due to insufficient statistics. The remaining points were binned logarithmically -- the first bin was derived by averaging the two remaining lowest frequency points, the second by binning the next four points, eight points, etc. This binning scheme gives equal weightings to equal decades in frequency (Edelson \& Nandra 1998), when, e.g., fitting a power-law relation to the power spectrum.

\begin{figure}
\centerline{\epsfxsize=5.0in\epsfbox{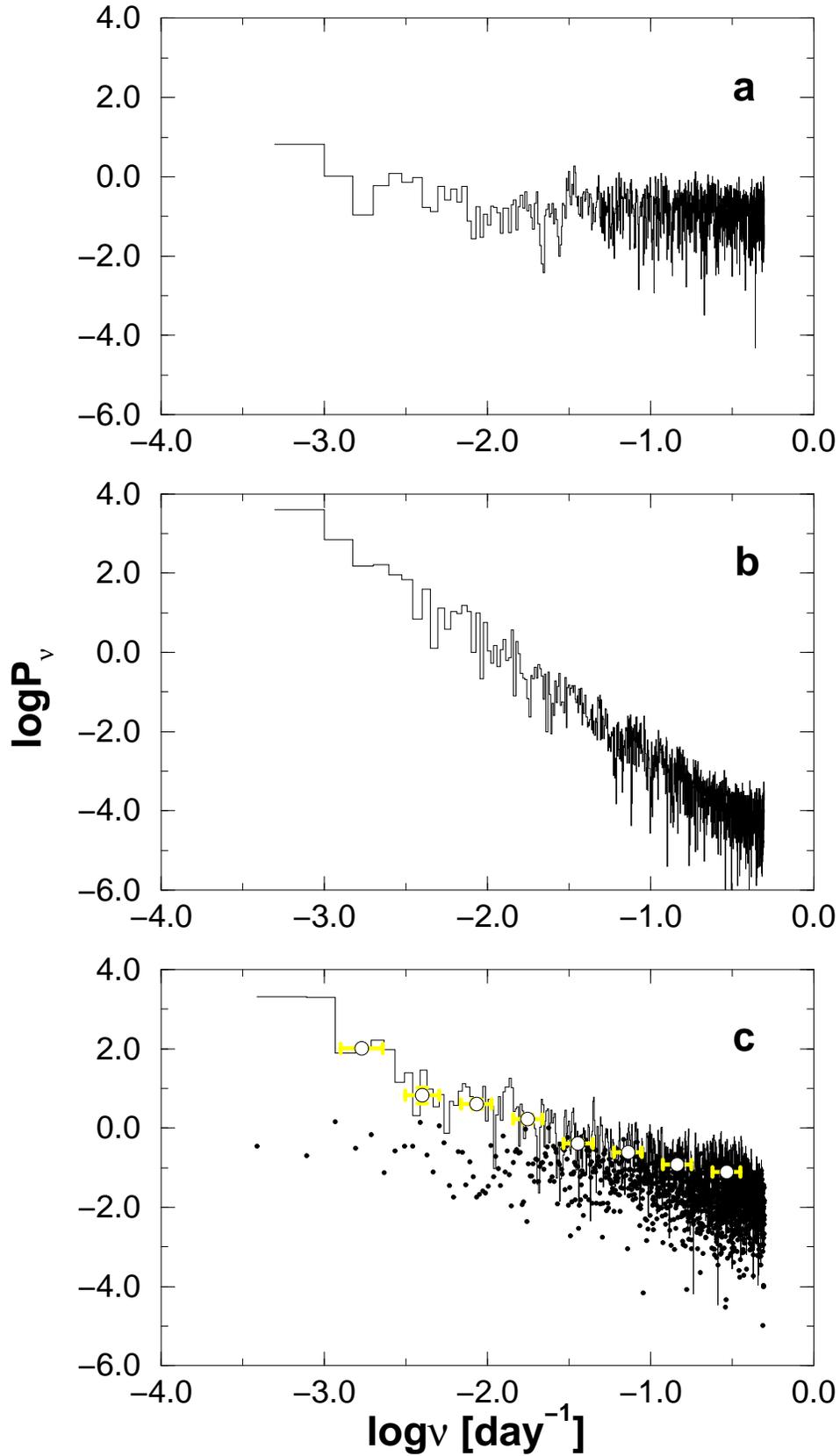}}
\caption{The $R$ band power spectrum of PG0804+762, in logarithmic units: a) Power spectrum obtained without any interpolation of the unevenly sampled data. b) Power spectrum after full interpolation of the lightcurve to an evenly-sampled grid. c) Power spectrum resulting from the partial interpolation scheme described in the text. The power spectrum of the window function is shown for comparison in the lower panel.}
\label{pg0804spec}
\end{figure}

As an example, we show in Figure \ref{pg0804spec} three power spectra of  PG0804+762. Figure \ref{pg0804spec}a is the power spectrum resulting from the uninterpolated lightcurve. Due to the suppression of the low frequencies, it appears almost like a white-noise spectrum. Figure \ref{pg0804spec}b is the power spectrum of the fully interpolated lightcurve. Here the higher frequencies are attenuated by the smoothing effect of the interpolation. Figure \ref{pg0804spec}c shows the power spectrum resulting from our partial interpolation scheme.
This power spectrum is similar to the ``fully-interpolated'' spectrum (Fig. \ref{pg0804spec}b) at low frequencies and to the ``uninterpolated'' spectrum (Fig. \ref{pg0804spec}a) at high frequencies, with a smooth transition between them.
In the lower panel we also show the power spectrum of the window function. It was obtained by sampling, with the same temporal pattern, white noise having our measurement errors and calculating the power spectrum with this algorithm.

To test the reliability of our procedure in reproducing a given power-law index, we have carried out Monte-Carlo simulations. We have generated artificial lightcurves by summing sine waves with amplitudes following a power-law relation and having random phases.
Multiple lightcurves were produced by drawing different random phases. We then sampled each lightcurve with the actual temporal pattern of one of the observed quasars selected at random, and measured the distribution of the power-spectrum indices produced by our partial-interpolation procedure.
We find from these simulations that the input index is best reproduced by fitting only frequencies $\leq 0.01$ day$^{-1}$.
The $1\sigma$ scatter about the index that is input to the Monte-Carlo simulations is typically $\pm 0.6$. The mean of the Monte-Carlo distribution of indices matches the input index when the denser sampling patterns (with more than 60 points) are used. For sparser sampling patterns the mean output index is systematically different from the input index. The power spectrum of a daily sampled lightcurve in our simulations reproduced the input index with an accuracy of $\sim\pm 0.1$. 

We have applied this algorithm to our data in the frequency range of 0.001 day$^{-1}\leq\nu\leq 0.01$ day$^{-1}$. We find that the quasars generally have power spectra in this range of frequencies, that can be fitted with a single power-law with $\chi^2$ values of unity order. Exceptions are PG1444+407, PG1415+451 ($B$ band), and PG2233+134 ($R$ band), whose power spectra show a break at $\sim 300$ days.
The index values for the $B$ and $R$ lightcurves are in the ranges $0.8<\gamma_B<3.3$, and $0.6<\gamma_R<3.1$, and both have means of $\gamma=2.0\pm 0.6$.
Based on the Monte-Carlo simulations, we therefore cannot distinguish the spectral indices of the individual PG quasars.
However, we can conclude that the quasars have a limited range of a power-law indices, $\ltorder\pm 0.6$.

{\bf Autocorrelation Analysis -}
The autocorrelation function (ACF) is defined by
\begin{equation}
ACF(\tau)\equiv\big< \big( m(t)-\langle m\rangle\big)\cdot\big( m(t+\tau)-\langle m\rangle\big)\big>,
\label{define_acf}
\end{equation}
where brackets denote a time average.
The ACF measures the correlation of the lightcurve with itself, shifted in time, as a function of the time lag $\tau$. If there is an underlying signal in the lightcurve, with a typical time scale, then the width of the ACF peak near zero time-lag will be proportional to this time scale.
Another function used in variability studies to estimate the variability at different time lags is the first-order structure function (SF; e.g., Trevese et al. 1994) defined by
\begin{equation}
SF(\tau)\equiv\sqrt{\big< \big( m(t+\tau)- m(t)\big)^2\big>}.
\label{SF}
\end{equation}
There is a simple relation between the ACF and the SF, 
\begin{equation}
SF^2(\tau)=2\left(V-ACF(\tau)\right),
\label{acf_sf}
\end{equation}
where $V$ is the variance of the lightcurve. We therefore perform only an ACF analysis on our lightcurves. 

\begin{figure}
\centerline{\epsfxsize=3.0in\epsfbox{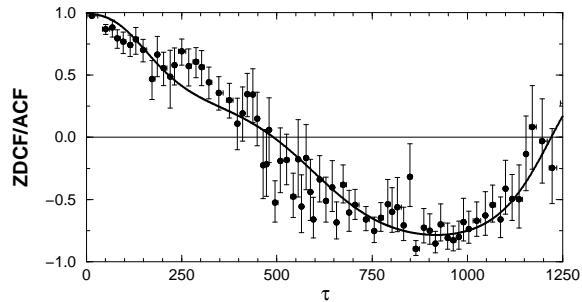}}
\caption{The ZDCF (discrete points) and its fit (fifth order polynomial least-squares fit, solid line) for the PG0804+762 $R$ band data.}
\label{pg0804}
\end{figure}

A ``correlation time scale'' can be defined by the width of the ACF peak. For a power-law-type power spectrum, that does not have a characteristic time scale, the power-law index and the $S/N$ determine the correlation time scale. In our analysis we choose the zero-crossing time of the ACF as the correlation time scale, which designates the time when the signal has entirely ``forgotten'' its past.
We have estimated the autocorrelation function (ACF) using the z-transformed discrete correlation function (ZDCF, Alexander 1997). Alexander (1997) has shown that this method is statistically robust even when applied to very sparsely and irregularly sampled lightcurves ($n_{obs}\geq 12$). The ZDCF was calculated for all of the lightcurves. We then used a least-squares procedure to fit a fifth-order polynomial to the ZDCF, with the constraint that $ACF(\tau=0)=1$. We used the ZDCF fit to evaluate the zero-crossing time in the observer's frame. 
As an example of the autocorrelation analysis, the ZDCF with its polynomial fit for PG0804+762 is shown in Figure \ref{pg0804}.
ACF time scales are listed in Table 3.

\vspace{1cm}

The statistics described above are listed in Table 3 in the following format:
\begin{figure*}
\includegraphics{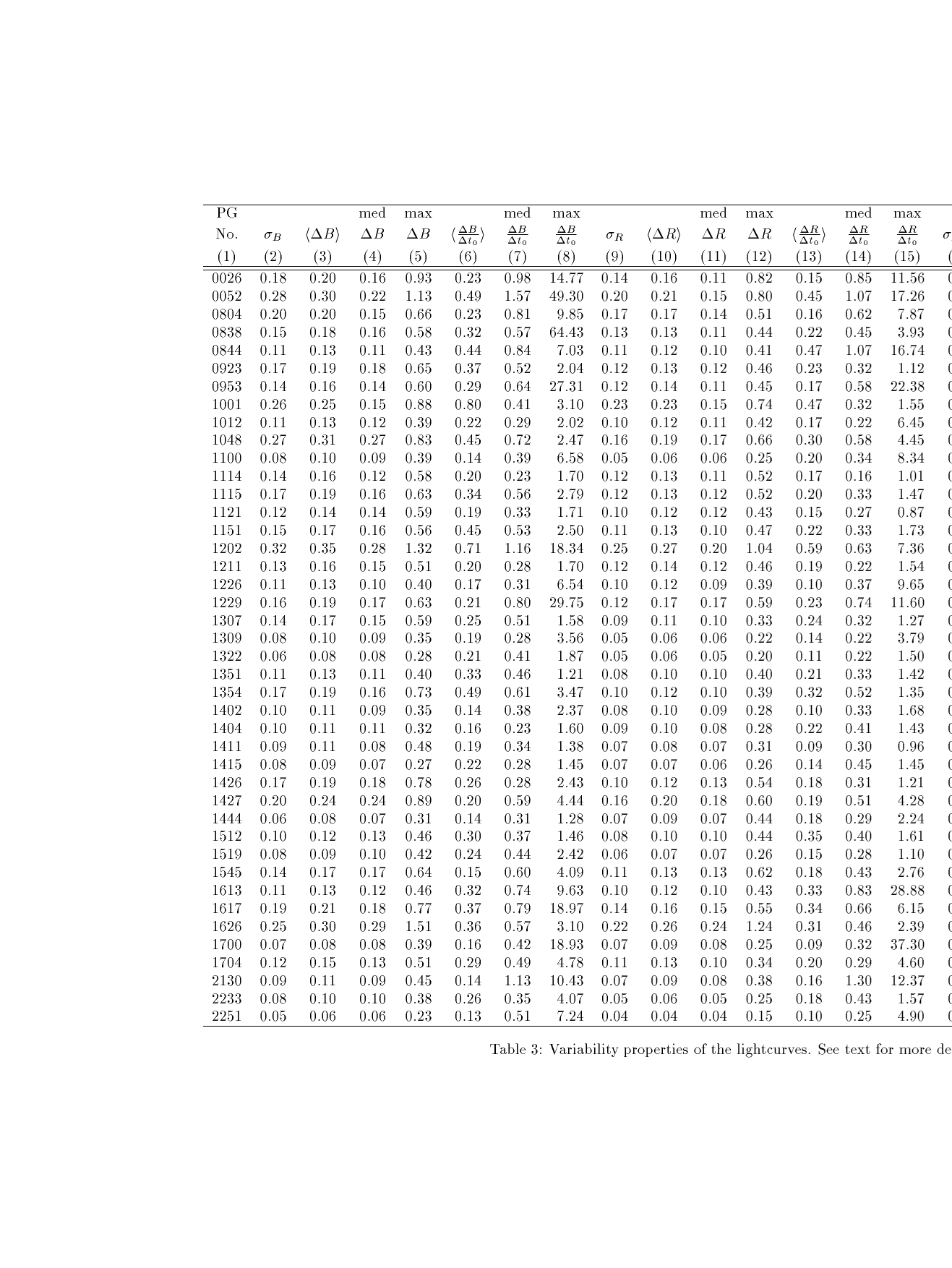}
\vspace{25cm}
\end{figure*}
\begin{table*}
\label{results}
\end{table*}
\addtocounter{table}{1}
\begin{itemize}
\item[]{\it Column (1)\/}---PG name (shortened).
\item[]{\it Column (2)\/}---Standard deviation of the lightcurves in the $B$ band.
\item[]{\it Columns (3)-(5)\/}---Mean, median,and maximal differences of the lightcurves in the $B$ band.
\item[]{\it Columns (6)-(8)\/}---Mean, median,and maximal rest-frame gradients of the lightcurves in the $B$ band.
\item[]{\it Columns (9)-(15)\/}---Same as columns (2)-(8), in the $R$ band.
\item[]{\it Columns (16)-(22)\/}---Same as columns (2)-(8), for the $B-R$ colour.
\item[]{\it Column (23)\/}---Net difference of the lightcurves, in the $B$ band.
\item[]{\it Column (24)\/}---Same as column (23), in the $R$ band.
\item[]{\it Column (25)\/}---Observer's frame ACF time scale, in the $B$ band, in days. A ``$\star$'' is marked for objects whose ACF did not reach zero correlation. These quasars were excluded from the correlation analysis.
\item[]{\it Column (26)\/}---Same as column (25), in the $R$ band.
\end{itemize}

{\bf Correlation Analysis -}
We have tested for the existence of correlations among all of the measured quantities described above. We have also searched for correlations with other parameters known for the PG sample (Kellerman et al. 1989; Boroson \& Green 1992; Laor et al. 1997) that are listed in Table 2. Two additional data sets were included in the correlation analysis -- the IR spectral indices from Neugebauer et al. (1987), and the polarization data from Berriman et al. (1990). Of special interest are the X-ray spectral index ($\alpha_x$) and the H$\beta$ FWHM, that were found by Fiore et al. (1998) to be correlated with the X-ray variability amplitude on short (2--20 days) time scales, but uncorrelated on longer time scales. We also sought correlations in the subclasses of the radio-loud quasars (RLQ, quasars with radio-to-optical flux ratio, $\rm{R}>10$) and the radio-quiet quasars (RQQ, $\rm{R}<10$) in our sample. There are 7 RLQ and 35 RQQ in our sample.

We chose Spearman's rank-correlation coefficient over Pearson's correlation coefficient because it tests for a general monotonic relation rather than only a linear one. However, in cases of bimodality Spearman's coefficient may fail to detect correlations (e.g., the variability-amplitude/luminosity correlation which we discuss below). Table 4 shows the probability matrix of the quasar properties, where each entry gives the probability that the data are drawn from two uncorrelated populations.

There are 322 correlations in Table 4, so naively one expects 16 spurious correlations with $P_r\leq 5\%$. We have found 51 correlations with probabilities $\leq 5\%$. However, only 11 of the correlations are between independent parameters. The remaining 40 are between dependent parameters ({\it apriori \/} known to be correlated) or degenerate parameters (functions of each other).
There are only two independent variability parameters -- the variability amplitude and the variability time scale.
This is because the variability amplitude parameters for each band are degenerate variables, since they are functions (although complicated) of each other, and the two bands are dependent. Therefore, all of the variability amplitude parameters should be counted only once. The variability time scales in the two bands are also dependent.
Similarly, the luminosities in both bands are dependent variables, and both are coupled to the redshift, since the PG sample is flux-limited (see below); the radio-loudness and the radio power are both coupled to the luminosity (Hooper et al. 1996, see below); the H$\beta$ FWHM and $\alpha_x$ are also correlated (Laor et al. 1997), and $\alpha_x$ and $\alpha_{ox}$ are dependent variables, too.
Therefore, there are 8 independent spectral properties. This gives a total of 16 {\it independent \/} correlations, for which 1.6, and 0.8 are expected to be spurious for $P_r\leq 10\%$ and $P_r\leq 5\%$, respectively. Thus, given that we find 11 independent correlations, we expect $\ltorder 1.1$ of them to be spurious.
\begin{figure*}
\includegraphics{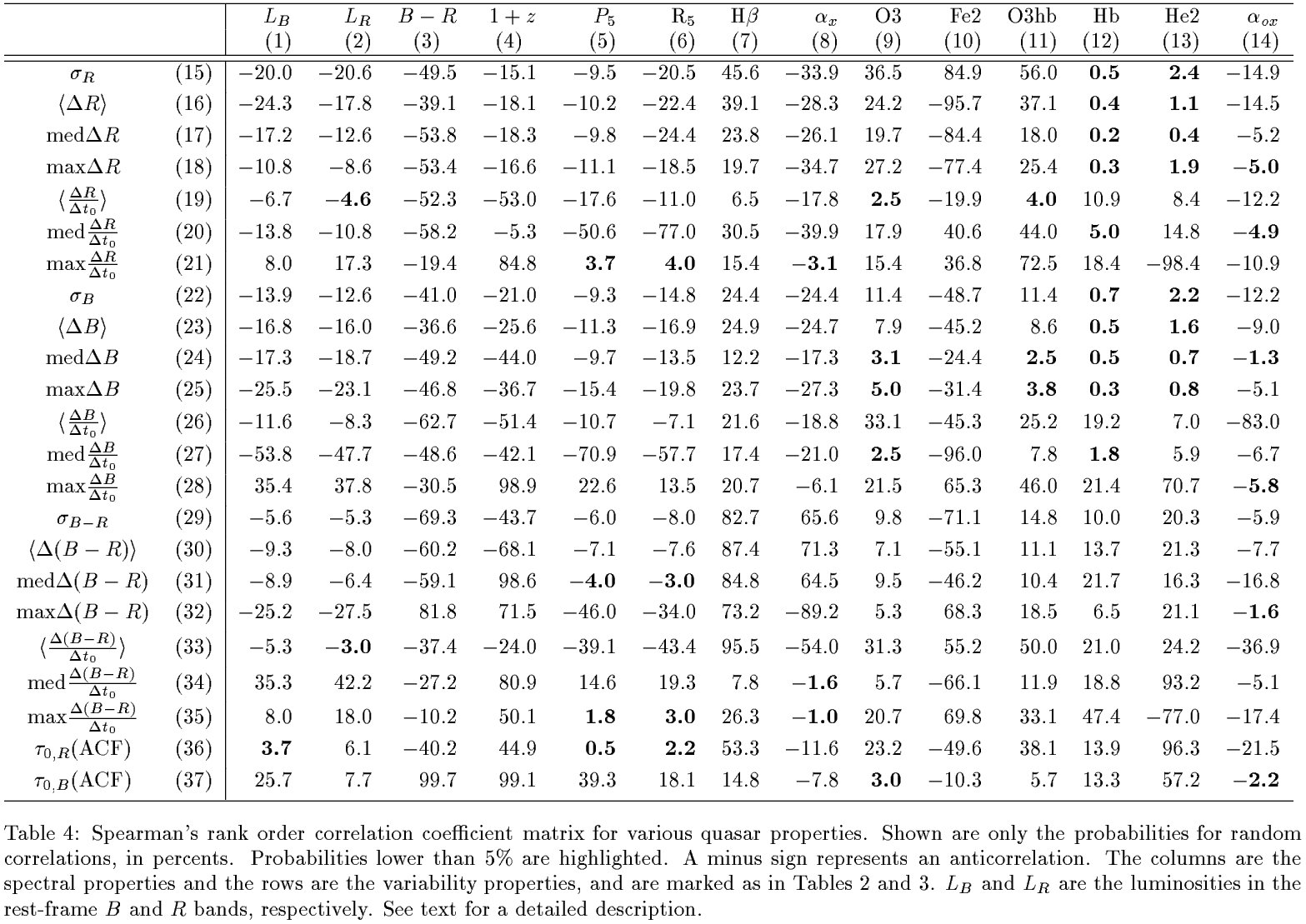}
\vspace{25cm}
\end{figure*}
\begin{table*}
\label{corrmatrix}
\end{table*}
\addtocounter{table}{1}We discuss the significant correlations listed in Table 4 in \S \ref{discussion}.

We have also tested for correlations between the change in the $B-R$ colour and the change in brightness in individual objects. It is established in the best-studied Seyfert galaxies that as the continuum source brightens, its UV-optical spectrum becomes harder, i.e., the variability amplitude increases with the emitted frequency (e.g., Maoz et al. 1993, Romano \& Peterson 1998). However, we have found only one work, which had demonstrated this behaviour in a sample of quasars (Cutri et al. 1985). We have calculated the correlations between the $B-R$ colour and both the $B$ and the $R$ magnitudes for each object.
The results are shown in Table \ref{colourmag}.
\begin{table}
\begin{tabular}{|c|r|c|r|c|} \hline
PG No. & $r_S(B)$ & $P_r(B)$ & $r_S(R)$ & $P_r(R)$ \\ \hline\hline
0026  & {\bf 0.81}  & {\bf 6.2$\times 10^{-18}$}  & {\bf 0.65}  & {\bf 6.2$\times 10^{-10}$} \\
0052  & {\bf 0.91}  & {\bf 7.6$\times 10^{-29}$}  & {\bf 0.78}  & {\bf 2.6$\times 10^{-16}$} \\
0804  & {\bf 0.75}  & {\bf 5.9$\times 10^{-17}$}  & {\bf 0.58}  & {\bf 3.6$\times 10^{-9}$} \\
0838  & {\bf 0.71}  & {\bf 1.6$\times 10^{-5}$}   &      0.29   &      1.3$\times 10^{-1}$  \\
0844  &      0.19   &      1.3$\times 10^{-1}$    &   $-$0.17   &      1.7$\times 10^{-1}$  \\
0923  & {\bf 0.91}  & {\bf 8.3$\times 10^{-10}$}  & {\bf 0.72}  & {\bf 7.1$\times 10^{-5}$} \\
0953  & {\bf 0.67}  & {\bf 6.0$\times 10^{-9}$}   & {\bf 0.37}  & {\bf 3.5$\times 10^{-3}$} \\
1001  & {\bf 0.47}  & {\bf 1.7$\times 10^{-2}$}   &      0.08   &      7.2$\times 10^{-1}$  \\
1012  &      0.33   &      1.2$\times 10^{-1}$    &   $-$0.21   &      3.5$\times 10^{-1}$  \\
1048  & {\bf 0.85}  & {\bf 1.8$\times 10^{-9}$}   & {\bf 0.57}  & {\bf 9.4$\times 10^{-4}$} \\
1100  & {\bf 0.87}  & {\bf 1.1$\times 10^{-15}$}  & {\bf 0.50}  & {\bf 3.2$\times 10^{-4}$} \\
1114  & {\bf 0.41}  & {\bf 4.4$\times 10^{-2}$}   &      0.06   &      7.8$\times 10^{-1}$  \\
1115  & {\bf 0.79}  & {\bf 2.4$\times 10^{-6}$}   & {\bf 0.50}  & {\bf 1.0$\times 10^{-2}$} \\
1121  & {\bf 0.57}  & {\bf 3.0$\times 10^{-3}$}   &      0.17   &      4.1$\times 10^{-1}$  \\
1151  & {\bf 0.78}  & {\bf 1.2$\times 10^{-5}$}   & {\bf 0.54}  & {\bf 8.3$\times 10^{-3}$} \\
1202  & {\bf 0.80}  & {\bf 3.4$\times 10^{-9}$}   & {\bf 0.66}  & {\bf 9.3$\times 10^{-6}$} \\
1211  & {\bf 0.54}  & {\bf 6.6$\times 10^{-3}$}   &      0.23   &      2.7$\times 10^{-1}$  \\
1226  & {\bf 0.56}  & {\bf 7.0$\times 10^{-5}$}   &      0.24   &      1.2$\times 10^{-1}$  \\
1229  & {\bf 0.35}  & {\bf 2.3$\times 10^{-2}$}   &{\bf $-$0.33}& {\bf 3.1$\times 10^{-2}$} \\
1307  & {\bf 0.78}  & {\bf 3.9$\times 10^{-7}$}   & {\bf 0.53}  & {\bf 2.4$\times 10^{-3}$} \\
1309  & {\bf 0.75}  & {\bf 2.3$\times 10^{-6}$}   &      0.21   &      2.7$\times 10^{-1}$  \\
1322  & {\bf 0.72}  & {\bf 2.2$\times 10^{-5}$}   &      0.37   &      5.5$\times 10^{-2}$  \\
1351  & {\bf 0.53}  & {\bf 9.4$\times 10^{-4}$}   &      0.26   &      1.3$\times 10^{-1}$  \\
1354  & {\bf 0.81}  & {\bf 1.1$\times 10^{-6}$}   & {\bf 0.60}  & {\bf 1.4$\times 10^{-3}$} \\
1402  & {\bf 0.58}  & {\bf 1.2$\times 10^{-3}$}   &      0.05   &      7.8$\times 10^{-1}$  \\
1404  & {\bf 0.71}  & {\bf 6.3$\times 10^{-5}$}   &      0.12   &      5.6$\times 10^{-1}$  \\
1411  & {\bf 0.67}  & {\bf 9.0$\times 10^{-5}$}   &      0.28   &      1.5$\times 10^{-1}$  \\
1415  & {\bf 0.49}  & {\bf 8.7$\times 10^{-3}$}   &   $-$0.09   &      6.5$\times 10^{-1}$  \\
1426  & {\bf 0.86}  & {\bf 3.8$\times 10^{-8}$}   & {\bf 0.69}  & {\bf 1.6$\times 10^{-4}$} \\
1427  & {\bf 0.66}  & {\bf 1.8$\times 10^{-4}$}   & {\bf 0.55}  & {\bf 2.9$\times 10^{-3}$} \\
1444  &      0.01   &      9.5$\times 10^{-1}$    &{\bf $-$0.42}& {\bf 3.3$\times 10^{-2}$}  \\
1512  & {\bf 0.64}  & {\bf 3.8$\times 10^{-5}$}   & {\bf 0.45}  & {\bf 7.1$\times 10^{-3}$} \\
1519  & {\bf 0.65}  & {\bf 5.1$\times 10^{-5}$}   &      0.20   &      2.6$\times 10^{-1}$  \\
1545  & {\bf 0.88}  & {\bf 3.1$\times 10^{-14}$}  & {\bf 0.72}  & {\bf 8.7$\times 10^{-8}$} \\
1613  & {\bf 0.37}  & {\bf 2.6$\times 10^{-3}$}   &   $-$0.04   &      7.3$\times 10^{-1}$  \\
1617  & {\bf 0.74}  & {\bf 9.0$\times 10^{-11}$}  & {\bf 0.53}  & {\bf 3.1$\times 10^{-5}$} \\
1626  & {\bf 0.77}  & {\bf 1.1$\times 10^{-6}$}   & {\bf 0.60}  & {\bf 6.2$\times 10^{-4}$} \\
1700  &      0.14   &      3.0$\times 10^{-1}$    &{\bf $-$0.31}& {\bf 2.2$\times 10^{-2}$} \\
1704  & {\bf 0.61}  & {\bf 3.0$\times 10^{-5}$}   & {\bf 0.34}  & {\bf 3.4$\times 10^{-2}$} \\
2130  & {\bf 0.48}  & {\bf 1.2$\times 10^{-5}$}   &      0.07   &      5.2$\times 10^{-1}$  \\
2233  & {\bf 0.90}  & {\bf 2.0$\times 10^{-12}$}  & {\bf 0.40}  & {\bf 2.4$\times 10^{-2}$} \\
2251  & {\bf 0.62}  & {\bf 1.7$\times 10^{-5}$}   &   $-$0.02   &      9.1$\times 10^{-1}$  \\
\hline
\end{tabular}
\caption{Spearman's correlation coefficients and probabilities for the $B-R$ colour changes $vs.$ the $B$ and $R$ magnitude changes for individual objects.}
\label{colourmag}
\end{table}

\subsection{Correlation Results}

\label{discussion}

In this section we review the most significant correlations detected in the data, as listed in Tables 4 and \ref{colourmag}. The section is divided on the basis of the main variability properties measured for the sample -- the correlation of changes in colour with changes in magnitude, variability amplitude correlations, variability time scale correlations, variability gradient correlations, and colour variability correlations. All of the probabilities referred to in this section are taken from these tables, unless otherwise stated.

\begin{figure}
\centerline{\epsfxsize=3.0in\epsfbox{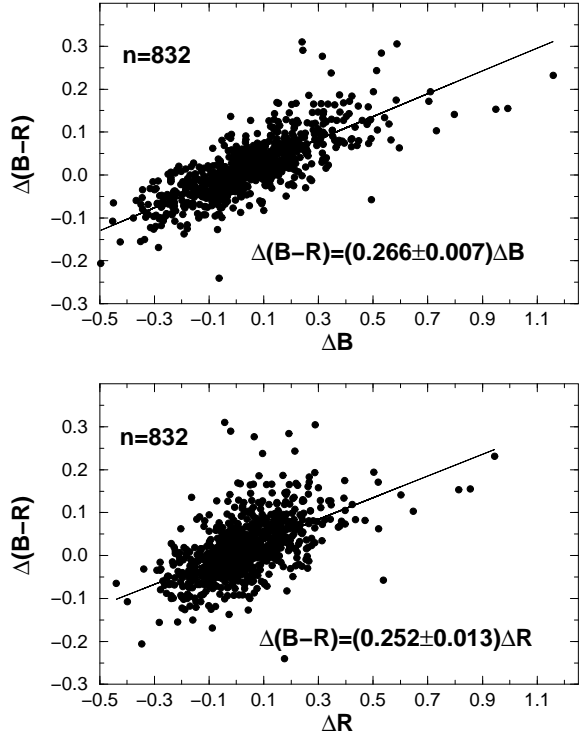}}
\caption{Deviations from the lightcurve means of the $B-R$ colour $vs.$ the deviations of the $B$ and $R$ magnitudes, for the data points of the 21 quasars that become bluer as they brighten. The number of data points and the best fit slopes for each band are specified.}
\label{colormag}
\end{figure}

{\bf Colour-Change Correlations -}
In Table \ref{colourmag} we have listed Spearman's correlation coefficients and probabilities for the correlations between the colour changes and the magnitude changes in the $B$ and $R$ bands for individual objects.
Thirty eight objects in Table \ref{colourmag} show significant correlations ($P_r\leq5\%$) of the colour changes with the changes in the $B$ magnitude, while only twenty three objects display this behaviour in the $R$ band. We suspect that the apparently-significant $\Delta(B-R)$ $vs.$ $\Delta B$ correlations of some of the quasars are due to the correlated errors between $\Delta(B-R)$ and $\Delta B$, while some significant $\Delta(B-R)$ $vs.$ $\Delta R$ correlations are likely lost due to the corresponding anticorrelated errors (note that PG 1229+204, PG 1444+407, and PG 1700+518 become redder when they brighten, probably due to this effect, because it is seen only in the $R$ band). In any case, it is clear that a large fraction of the PG quasars become bluer as they brighten. In order to quantify this behaviour, we have divided the sample into a subgroup (1) of the 21 quasars which get bluer as they brighten ($P_r(R)\leq 5.5\%$), and a subgroup (2) of the 21 quasars which do not. Figure \ref{colormag} shows the deviations from the lightcurve means of the $B-R$ colour $vs.$ the deviations of the $B$ and the $R$ magnitudes, for the data points of subgroup 1. 
 This figure manifests very significant correlations. The best linear fits are marked in the figure.

\begin{figure}
\centerline{\epsfxsize=3.0in\epsfbox{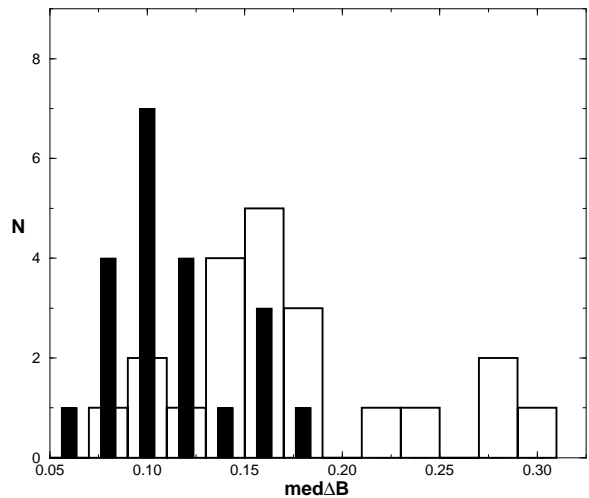}}
\caption{Distributions of the median amplitude difference in the $B$ band. Empty bars are objects which become bluer when they brighten, and filled bars are objects which do not exhibit a correlation between their changes in colour and brightness.} 
\label{bluish}
\end{figure}

To investigate if any of the quasar parameters is driving this division of our sample, we checked for correlations between the various quasar parameters, referred to in the preceding sections, and the probabilities $P_r(R)$ in Table \ref{colourmag}. We find that the quasars with the larger variability amplitudes are those that tend to become bluer when brightening in $R$. Figure \ref{bluish} shows the histograms of the median difference in the $B$ band, which represents the variability amplitude, with the two subgroups distinguished. A Student's t-test on the means of the two subgroups shows them to be different at a significance level $\geq 99\%$.  From the histograms it is evident that for $\Delta B\gtorder 0.2$ mag, all objects become significantly bluer when they brighten. The range $0.1\ltorder\Delta B\ltorder0.2$ mag is occupied by both populations. For $\Delta B\ltorder0.1$ mag, the uncertainty of the colour measurements becomes dominant [$\Delta(B-R)=(0.266\pm 0.007)\Delta B$, Figure \ref{colormag}], and a definitive statement about the colour variations cannot be made.
Some role in this effect may be played by the colours and magnitudes of the host galaxies, but investigating this will require improved imaging data for the hosts of this sample.

\begin{figure}
\centerline{\epsfxsize=4.0in\epsfbox{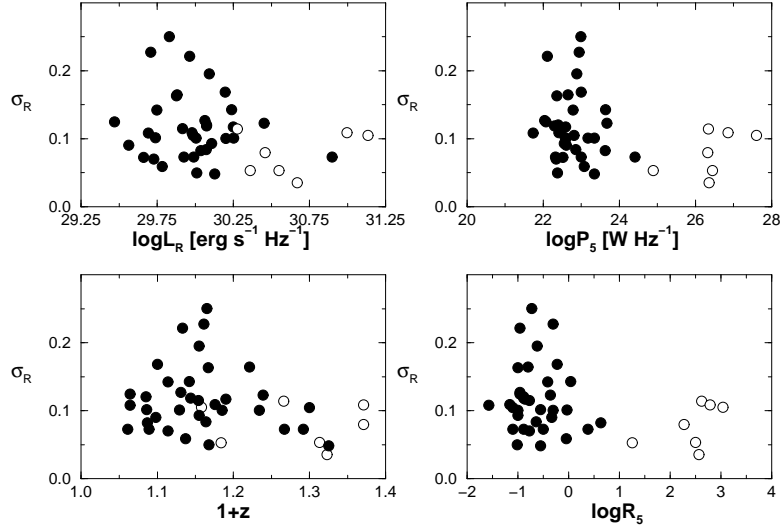}}
\caption{$R$ band standard deviation $vs.$ rest-frame $R$ band luminosity ($L_R$), redshift ($1+z$), total radio power at 5 GHz ($P_5$), and the radio-loudness ($R_5$). Filled circles are radio-quiet objects and empty circles are radio-loud objects.}
\label{sigmaR-LzPR}
\end{figure}

{\bf Variability Amplitude Correlations -}
We find that the variability amplitude in its various forms is weakly anticorrelated with the source optical luminosity with probabilities of $P_r=9-20\%$ for the $R$ band and of $P_r=13-26\%$ for the $B$ band. This is in accord with previous claims by most of the authors mentioned in Table \ref{former_works} (Pica \& Smith 1983; Trevese et al. 1994; Hook et al. 1994; Hawkins 1996; Cristiani et al. 1997). The significance of this correlation is weaker than that found in the study of Pica \& Smith (1983) -- $P_r=7\%$ for RQQ, and weaker than found by Hook et al. (1994) and Cristiani et al. (1997), which is $P_r\leq 1\%$. The same weak trend exists also between the variability amplitude and the radio power (at 5 GHz), at a significance level of $P_r=10-11\%$ (However, Impey et al. (1991) have found that for RQQ the variability amplitude is correlated with the radio power). In Figure \ref{sigmaR-LzPR} we show the $R$ band standard deviation $vs.$ luminosity, redshift, radio power, and radio-loudness.
 The lower mean variability amplitude of the RLQ is clear, even though there is no significant monotonic trend according to Spearman's rank test ($P_r=14-24\%$).
However, the quasars with the highest optical luminosity in the PG sample are RLQ (as already noted by Hooper et al. 1996; our data give a correlation of $r_S=0.512$, $P_r=5.3\times 10^{-4}$ between $L_R$ and $R$), so this may be a luminosity effect. From the $\sigma_R$ $vs.$ $\log{L_R}$ plot, no significant difference exists in variability amplitude between objects in the RLQ and the RQQ subgroups that have comparable luminosities ($\sigma_{\rm RL}=0.08\pm0.03$ and $\sigma_{\rm RQ}=0.10\pm0.04$, and are indistinguishable at the 80\% significance level according to Student's t-test). Hence we cannot say whether the low amplitude of the RLQ is the result of their high optical luminosity or their radio power, and there is no evidence for the RLQ having {\it larger\/} variability amplitude than the RQQ, in agreement with the findings of Pica \& Smith (1983) and Borgeest \& Schramm (1994). 

We find no significant correlation between the variability amplitude parameters and the source redshift. However (see Figure \ref{sigmaR-LzPR}), here too there may be a bimodality, with $z<0.2$ objects having larger $\sigma_R$ than $z>0.2$ objects ($\sigma_{z<0.2}=0.12\pm0.05$ and $\sigma_{z>0.2}=0.09\pm0.04$, which are different at the $95\%$ significance level according to Student's t-test).
We note also that in the lower redshift bin ($z<0.2$) the trend of the variability amplitude-redshift correlation is opposite to the trend in the higher redshift bin ($z>0.2$). The PG sample suffers from a well-known luminosity-redshift bias, as do other flux-limited samples (Schmidt \& Green 1983). Our data give a correlation coefficient $r_S=0.608$, $P_r=1.9\times 10^{-5}$, between the rest-frame $R$ band luminosity and the redshift. Thus, the correlations of variability amplitude with luminosity and redshift cannot be disentangled in our sample.

\begin{figure}
\centerline{\epsfxsize=3.0in\epsfbox{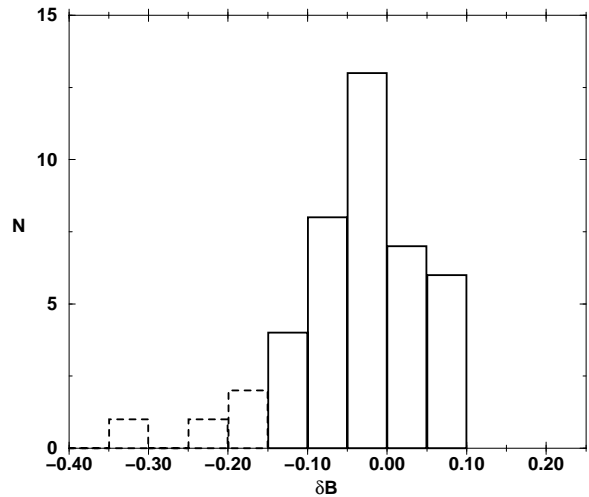}}
\caption{$B$ band net difference histogram. Dashed-lined bars correspond to the four objects (PG 0026+129, PG 1048+342, PG 1427+480, and PG 1626+554) exhibiting long-term monotonic brightening.}
\label{upsndowns}
\end{figure}

The variability amplitude is anticorrelated with the net difference (equation \ref{netdiff}) with significance $\gtorder 99\%$ (not shown in Table 4). This means that objects that show larger variations exhibit asymmetry (negative $\delta B$) in their variations -- on average the brightening takes longer then the fading. However, this trend is dominated by four objects (PG 0026+129, PG 1048+342, PG 1427+480, and PG 1626+554) whose lightcurves exhibit long term monotonic variations and therefore have large $\delta B$ and $\delta R$. Excluding them from the correlation analysis reduces the above significance to $95-99\%$. The net difference histogram is shown in Figure \ref{upsndowns}, with the four monotonically rising objects highlighted.
When these four objects are excluded, the mean of the distribution is $\delta B=-0.02$, which is consistent with zero according to Student's t-test.

\begin{figure}
\centerline{\epsfxsize=3.0in\epsfbox{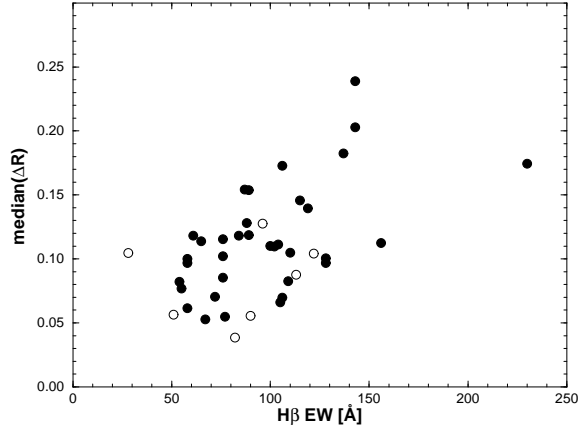}}
\caption{The median difference in the $R$ band $vs.$ H$\beta$ EW. Filled circles are radio-quiet objects and empty circles are radio-loud objects.}
\label{dR-Hb}
\end{figure}

A strong result of our study is a correlation of the variability amplitude with the H$\beta$ EW. It emerges from all of our variability amplitude test parameters, and in both bands (see Table 4). The average difference in the $R$ band is plotted $vs.$ H$\beta$ EW in Figure \ref{dR-Hb}.
A possible, but unlikely, explanation for this trend is the emission-line contribution to the broad-band fluxes. In objects with large emission-line EW, small intrinsic optical continuum variability may be accompanied by larger variations in the ionizing continuum, producing large variations in the optical emission lines (after a delay of several months, due to the light-travel time across the broad-line region; Kaspi et al. 1996).
The delayed and smoothed response could also explain the larger ACF time scales in quasars with larger EW (see below). However, in all AGN that have been monitored spectroscopically to date, the fractional amplitude of the emission-line variations is always considerably {\it smaller\/} than that of the optical continuum variations, once known constant components such as narrow lines and galaxy light have been accounted for (e.g., Maoz 1997).
Spectrophotometry of several of the PG quasars suggests this behaviour applies to this sample as well (Maoz et al. 1993; Kaspi et al. 1996). Spectrophotometric monitoring data for a subsample of the quasars may allow us to understand this correlation (Kaspi et al. 1999).
The correlation with the variability amplitude is also detected for HeII $\lambda 4686$ EW, [OIII] $\lambda 5007$ EW, and [OIII] $\lambda 5007$-to-H$\beta$ peak intensities ratio ($B$ band only), though at weaker significance levels (see Table 4).
Apparently, this trend is not a combination of the variability amplitude--luminosity anticorrelation and the Baldwin effect (${\rm EW}\propto L^{-\delta}$ for various lines; Baldwin 1977, Osmer et al. 1988), since H$\beta$ is one line that does not exhibit the Baldwin effect (Binette et al. 1993).

\begin{figure}
\centerline{\epsfxsize=3.0in\epsfbox{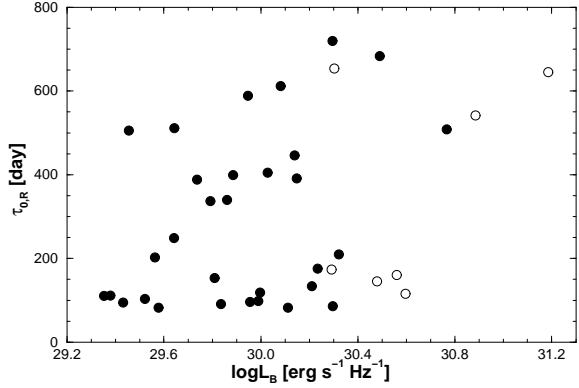}}
\caption{Rest-frame ACF time scale in the $R$ band $vs.$ the rest-frame $B$ band luminosity. Filled symbols are radio-quiet objects and empty symbols are radio-loud objects.}
\label{tau-L}
\end{figure}

{\bf Variability Time Scale Correlations -}
We find that the rest-frame ACF variability time scale is correlated with the source luminosity, but not correlated with redshift. The correlation with the luminosity is opposite to the ones found by Trevese et al. (1994) and Netzer et al. (1996). The strongest correlation ($P_r=2.1\%$) is between the rest-frame ACF time scale in the $R$ band and the rest-frame $B$ band luminosity, which are plotted in Figure \ref{tau-L}.
Four quasars (all RQQ), whose ACF did not reach zero correlation (see Table 3), were excluded from the calculations.
 
\begin{figure}
\centerline{\epsfxsize=3.0in\epsfbox{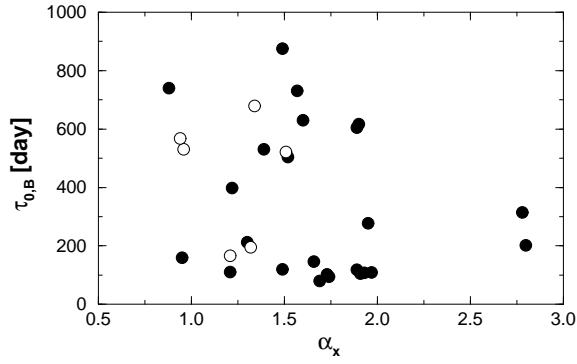}}
\caption{Rest-frame ACF time scale in the $B$ band $vs.$ $\alpha_x$. Filled symbols are radio-quiet objects and empty symbols are radio-loud objects.}
\label{tax}
\end{figure}

A weak anticorrelation exists between the $B$ band rest-frame ACF time scale and the X-ray spectral index, $\alpha_x$ ($P_r=7.8\%$), shown in Figure \ref{tax}.

\begin{figure}
\centerline{\epsfxsize=3.0in\epsfbox{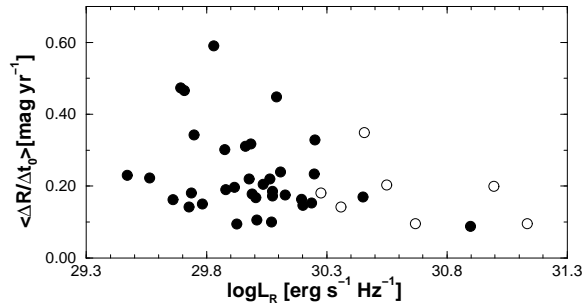}}
\caption{Mean rest-frame gradient in magnitudes per year $vs.$ monochromatic luminosity, $L_R$. Filled circles are radio-quiet objects and empty circles are radio-loud objects.}
\label{gradR-Lz}
\end{figure}

{\bf Variability Gradient Correlations -}
The median and the mean rest-frame gradients are anticorrelated with luminosity.
This is shown graphically in Figure \ref{gradR-Lz}.

We detect a few significant correlations between the rest-frame variability gradient and a few parameters, which seem peculiar, in that related measures, e.g., the mean and median gradients, show no comparable correlations. For example, the mean $R$ band gradient and the [OIII] $\lambda 5007$ to H$\beta$ peak intensities ratio are significantly correlated ($P_r=4.0\%$) but the median $R$ band gradient is not correlated with that parameter.

\begin{figure}
\centerline{\epsfxsize=3.0in\epsfbox{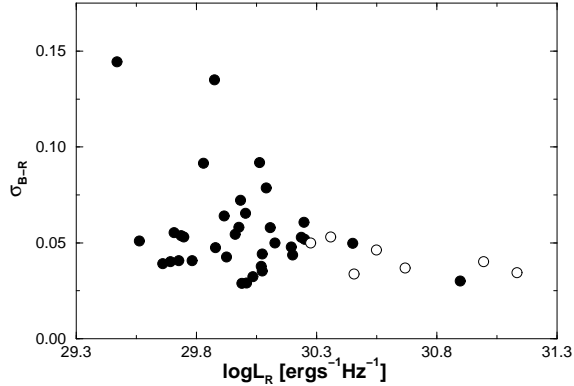}}
\caption{Relative rms amplitude of the $B-R$ colour $vs.$ rest frame $R$ band luminosity, $L_R$. Filled circles are radio-quiet objects and empty circles are radio-loud objects.}
\label{sigBR-L}
\end{figure}

{\bf Colour Variability Correlations -}
We have also tested for correlations of the quasar parameters with the $B-R$ colour variation properties (see Tables 3 and 4). All of the parameters used to estimate the colour variability amplitude show significant anticorrelations with the optical luminosities, and some of them show anticorrelations with the radio power and the radio-loudness. The anticorrelations with the radio power may be due to the radio-loudness--luminosity trend combined with the colour-luminosity anticorrelation which is exhibited by the sample (Mushotzky \& Wandel 1989). In Figure \ref{sigBR-L} we show the $B-R$ rms amplitude $vs.$ the $R$ band rest-frame luminosity ($P_r=1.3\%$).

Less luminous objects tend to have larger colour gradients, as is evident from the colour mean gradient-luminosity anticorrelation ($P_r\leq 2\%$) seen in Table 4. The maximal colour gradient is correlated with the radio-loudness ($P_r=3.0\%$) but not with the optical luminosity. These correlations suggest that objects that are more radio-loud have colour lightcurves with flickering activity, but the total variance of their colour lightcurves is lower.

\vspace{0.5cm}

We have checked for correlations between the variability parameters and two additional data-sets -- the IR spectral indices (Neugebauer et al. 1987) and the polarization levels (Berriman et al. 1990) of the PG quasars, but no significant correlations were found. The PG quasars have low polarizations and very low polarization variability. The polarization is mostly due to scattering, as opposed to BL Lac objects, where synchrotron emission is responsible for it.

\subsection{Comparison With Other Studies}

\label{comparison}

As already outlined in \S \ref{introduction}, previous studies of quasar variability have often reached conflicting results.
For example, among the fourteen studies listed in Table \ref{former_works}, six found an anticorrelation between the variability amplitude and the source luminosity, and five did not detect such a correlation. The relation between the variability amplitude and the source redshift is even more confounding -- there are three positive correlations, one anticorrelation, and seven studies that did not detect any correlation.

In the present work we have detected a weak anticorrelation between the variability amplitude and the source luminosity, which agrees with most of the studies listed in Table~\ref{former_works}. Taking into account only studies of samples with luminosities and redshifts comparable to ours (B\`{o}noli et al. 1979; Giallongo et al. 1991; Cimatti et al. 1993; Trevese et al. 1994), our work agrees with the minority of Trevese et al. (1994).
No significant correlation was detected in our data between the variability amplitude and the source redshift, in accord with most of the previous studies. However, the sources redshifts of the our sample are in a rather narrow range ($z<0.4$) compared with other studies.
In any case, we find that these trends are weak and have intrinsically large scatter, as evidenced by the fact that they remain weak even when the observational noise is minimized.

\section{Conclusions}

\label{conclusions}

We have presented the results of a seven-year monitoring program of a sub-sample of 42 quasars from the PG sample, and studied their variability properties in the $B$ and $R$ bands. Our main findings are:
\begin{itemize}
\item[1.] All 42 objects have varied during the seven years with $5\%<\sigma_B<34\%$ and $4\%<\sigma_R<26\%$. The rms amplitudes and colour variation for the entire sample are $\sigma_B=14\%$, $\sigma_R=12\%$, and $\sigma_{B-R}=5\%$.
\item[2.] The power spectra of most of the objects have a power-law shape ($P_{\nu}\propto \nu^{-\gamma}$) on time scales of 100 to 1000 days. The mean and rms values for the entire sample are $\gamma=2.0\pm 0.6$, and this spread is comparable to our uncertainties in determining $\gamma$.
\item[3.] Our data reproduce the previously-reported anticorrelation between the variability amplitude and the luminosity, but this trend is weak. A weak anticorrelation between the variability amplitude and the redshift is also detected but may be the result of the luminosity-redshift bias of the PG sample.
\item[4.] Larger variations tend to be mildly asymmetric -- the brightening phase is longer than the fading phase.
\item[5.] The ACF time scale is correlated with luminosity. More luminous quasars thus exhibit variations with longer time scales.
\item[6.] The spectra of about half of the quasars become harder (bluer) when they brighten, and the colour changes of the rest are not correlated with brightness changes. The quasars that become bluer tend to be those that are most variable.
\item[7.] The RLQ tend to have lower variability amplitudes than the RQQ, but this may be due to the fact that the RLQ in this sample are more luminous than the RQQ.
\item[8.] Quasars that are more luminous and more radio-loud have smaller colour variations on average, but RLQ tend to exhibit flickering in their $B-R$ lightcurves.
\item[9.] The median rest-frame variability gradient is anticorrelated with luminosity and with redshift. In other words, variations are less abrupt at higher luminosities and higher redshift. We are not able to disentangle these anticorrelations from the luminosity-redshift bias.
\item[10.] Objects with larger equivalent widths of the H$\beta$, [OIII] $\lambda 5007$, HeII $\lambda 4686$ emission lines have larger variability amplitudes.
\end{itemize}

\vspace{0.5cm}

The overall picture emerging from these results is that the spectral and variability parameters which describe the PG quasars have two extremes. The higher-luminosity quasars, which are also radio-loud, bluer, and at higher redshift, exhibit lower amplitudes of variations with longer time scales, and their colour lightcurves exhibit flickering activity. At the other extreme, the lower-luminosity quasars, which in the PG sample are radio-quiet and at lower redshift, have a wide range of variability amplitudes and time scales but, on average, vary with larger amplitudes.

We point out, however, that the trends we have identified are weak and the scatter is always much greater than the trend itself. This holds even for the strongest trend we have found, between variation amplitude and emission-line EW. The only trend with relatively little scatter is the tendency of about half of the quasars to become bluer when they brighten and redder when they fade.

A main conclusion of this work is that none of the quasar parameters that we have examined are very reliable predictors of variability properties, even when observational noise (photometric errors, sampling frequency, and sample selection) are low. Future studies of other samples and other parameters may reveal improved variability predictors.

The mechanism behind quasar variability is unknown. Some models have been proposed to explain it, but none of them comprehensively account for the variability phenomenon and the entire range of other quasars properties. Currently debated models include models, which suggest that the variability is caused by instabilities in an accretion disk (e.g., Mineshige et al. 1994a,b; Bao \& Abramowicz 1996; Kawaguchi et al. 1998), starburst models, which argue that quasar variability is formed by superposing the independent signals from many supernova explosions (e.g., Cid Fernandes et al. 1996; Aretxaga et al. 1997), and microlensing models, which posit that much of the observed variability could be caused by external agents, such as intervening lenses (e.g., Hawkins 1996; see however, Alexander 1995). We defer to a future paper a detailed comparison of our results to the predictions of these and other models.  

\section*{Acknowledgements}

We thank Buell Jannuzi for contributing to the observations at Steward Observatory, and for valuable comments. 
Tal Alexander, Ari Laor, Tsevi Mazeh, Rudy Schild, T. Wang, and the anonymous referee are acknowledged for their useful input. We are grateful to Sammy Ben-Guigui, John Dan, Peter Ibbetson, Ezra Mashal, and Haim Mendelson of the Wise Observatory staff, and to Anna Heller for their assistance with the observations. This work was supported by grants from the Israel Science Foundation.

\bsp

\label{lastpage}

\end{document}